\documentclass[aps,prb,twocolumn,floatfix,showpacs]{revtex4-2}
\usepackage{amsmath,amssymb,graphicx,bm,color}
\def\veck{\mathbf k}
\def\vecq{\mathbf q}
\def\vecp{\mathbf p}
\def\vecQ{\mathbf Q}

\def\be{\begin{equation}}
\def\ee{\end{equation}}

\begin{document}
\title{Failure of the Baym-Kadanoff construction to match consistently quantum dynamics with thermodynamic critical behavior}

\author{V\'aclav  Jani\v{s}}  
\email{janis@fzu.cz}

\affiliation{Institute of Physics, The Czech Academy of Sciences, Na Slovance 2, CZ-18200 Praha  8,  Czech Republic}

 \author{\v{S}imon Kos} 
 
 \affiliation{Department of Physics and NTIS - European Centre of Excellence, University of West Bohemia, Univerzitní 8, CZ-301 00 Plzeň, Czech Republic}
 
 \author{Vladislav Pokorn\'y}

\affiliation{Institute of Physics, The Czech Academy of Sciences, Na Slovance 2, CZ-18200 Praha  8,  Czech Republic}

\date{\today}

%\maketitle

\begin{abstract}
We disclose a serious deficiency of the Baym-Kadanoff construction of thermodynamically consistent conserving approximations. There are two vertices in this scheme: dynamical and conserving.  The divergence of each indicates a phase instability. We show that each leads to incomplete and qualitatively different behavior at different critical points. The diagrammatically controlled dynamical vertex from the Schwinger-Dyson equation does not obey the Ward identity and cannot be continued beyond its singularity. The standardly used dynamical vertex alone cannot, hence, conclusively decide about the stability of the high-temperature phase. On the other hand, the divergence in the conserving vertex, obeying the conservation laws, does not invoke critical behavior of the spectral function and the specific heat.  Morerover, the critical behavior of the conserving vertex may become spurious in low-dimensional systems. Consequently, the description of the critical behavior of correlated electrons becomes consistent and reliable only if the fluctuations of the order parameter in the conserving vertex lead to a divergence coinciding with that of the dynamical one.        
\end{abstract}
%\pacs{72.15.Qm, 75.20.Hr}

\maketitle %newpage
\section{Introduction}

When the electron interaction energy is comparable with the kinetic energy, the impact of electron correlations cannot be treated perturbatively. Quantum dynamical fluctuations overtake control of the low-temperature behavior of strongly interacting electron systems. Consequently, critical phenomena and cooperative behavior with diverging response functions are expected. A reliable description of quantum dynamics with thermodynamic criticality is essential to interpret experimental findings and to design and predict the behavior of materials with the desired properties.     

Unfortunately, exact solutions of strongly correlated electron systems exist only for particular models and specific limiting cases. They are, nevertheless, paradigms of a consistent description of quantum criticality. The most important exact solutions of the low-temperature critical behavior are those of the single-impurity Anderson model (SIAM) \cite{Wiegmann:1983aa,Tsvelick:1983aa}, the Kondo model \cite{Wilson:1975aa,Andrei:1983aa}, and the Hubbard model in one spatial dimension \cite{Lieb:1968aa,Essler:2005aa}. In particular, the exact solution of the charge and spin symmetric state of the SIAM  offers a complete description of the low-temperature quantum criticality with the critical point at $T=0$ and $U=\infty$. A dimensionless Kondo scale $a$ controls the critical asymptotics at $U\to\infty$ with $\ln a = - \pi^{2}U \rho_{0}/8$, where $\rho_{0}$ is the density of states at the Fermi energy. This scale, defining a distance to the critical point, can be obtained from three functions: the inverse zero-temperature magnetic susceptibility $\chi$, the linear coefficient of the specific heat $\gamma$, and the width of the quasiparticle peak of the spectral function \cite{Hewson:1993aa}. Any consistent theory of critical behavior must deliver the same Kondo scale, at least qualitatively, whatever definition we use. In particular, the Wilson ratio must be restricted as $R= \chi/\gamma\in (1,2)$. 
 %Here, $\chi$ is the zero-temperature value of the magnetic susceptibility, and $\gamma$ is the linear coefficient of the specific heat.              

%Unfortunately, exact solutions are unavailable for most models of strongly correlated electron systems in real solids. That is why approximate schemes must be applied. 
With the lack of exact solutions, one needs a way to construct consistent approximations. Baym and Kadanoff set a framework for deriving thermodynamically consistent conserving approximations encompassing quantum dynamical fluctuations non-perturbatively \cite{Baym:1961aa,Baym:1962aa}. The construction is based on the Luttinger-Ward functional, from which both one-particle self-energy and two-particle-irreducible vertices can be derived via functional derivatives. The Baym-Kadanoff scheme is, however,  self-consistent only for the one-particle functions; no two-particle vertices explicitly enter the generating Lutinger-Ward functional. A two-particle self-consistency was added to the Baym-Kadanoff scheme by De Dominicis and Martin by including two-particle vertices \cite{DeDominicis:1963aa,DeDominicis:1964aa,DeDominicis:1964ab}. 

Dynamical fluctuations at intermediate coupling and at low temperatures drive the system to criticality with divergent two-particle vertices. Critical behavior in quantum many-body systems can be reached either by enhanced quantum fluctuations due to the non-commutativity of the operators of kinetic energy and interaction or by breaking the system's linear response to external forces.
Quantum dynamics is contained in the self-energy determined from the Schwinger-Dyson equation. Quantum fluctuations enter the Schwinger-Dyson equation via a dynamical two-particle vertex. Thermodynamic critical behavior is determined from the divergence of response functions and is accompanied by the emergence of symmetry-breaking order parameters. Response functions must be determined from two-particle vertices derived from the self-energy via a functional Ward identity and a Bethe-Salpeter equation to match the emergence of order parameters with the divergence in response functions \cite{Baym:1961aa}. 

The problem is that the two-particle vertex from the Schwinger-Dyson equation does not obey the Ward identity in any known approximate solution, and its divergence, hence, cannot be attributed to a thermodynamic critical behavior of conserving response functions \cite{Janis:1998aa,Janis:2017aa}.  
We, hence, have two ways to identify critical behavior in the Baym-Kadanoff construction. Quantum dynamical criticality is deduced from the divergence in the dynamical vertex entering the Schwinger-Dyson equation, and the thermodynamic one from the diverging conserving vertex leading to response functions satisfying macroscopic conservation laws. The real critical behavior is unique, and so the dynamical and conserving vertices must be identical in the exact solution to the quantum many-body problem \cite{Schwinger:1949aa,Martin:1959aa}. 
The existing two ways to match the single self-energy and the two-particle vertex in systems with correlated fermions make the definition of critical behavior in the Baym-Kadanoff construction ambiguous. Although this dichotomy was disclosed early, it has not yet been considered a severe problem. 
 
The two two-particle vertices derived from a single self-energy do not presently lead to the same critical behavior in any approximate scheme.  We recently suggested an alternative approach to resolve this dichotomy in the theoretical description of critical behavior.  We used a single two-particle irreducible vertex from the diverging Bethe-Salpeter equation to generate approximations with uniquely defined critical points \cite{Janis:2019aa}.  Consequently, we obtained two self-energies for the given two-particle vertex. We separated the self-energy with even and odd symmetry with respect to the symmetry breaking field determining the critical point in the Bethe-Salpeter equation. The even self-energy obeys the dynamical Schwinger-Dyson equation, and the odd one, generalizing the order parameter, is matched with a two-particle-irreducible vertex via a linearized Ward identity. We obtained a unique critical behavior, but a difference between the dynamical and conserving vertices still remained.

The aim of this paper is to demonstrate that the existence of two divergent vertices in approximate theories is a severe problem hindering the application of standard tools to describe critical behavior. The divergence in the Schwinger-Dyson equation and the divergence of the response functions lead to critical behavior at different critical points and set incompatible criteria for the stability of thermodynamic phases. Moreover, neither of the divergences individually determines the full and thermodynamically consistent critical behavior. The divergence in the Schwinger-Dyson equation is responsible for the critical behavior of the spectral function and the density of states, while the divergence of the response functions leads to the emergence of order parameters. The divergence-free dynamical vertex from the Schwinger-Dyson equation cannot then be used as a criterion for the stability of thermodynamic phases unless confirmed by a non-diverging conserving vertex. 

The layout of the paper is as follows. We introduce the thermodynamic functional generating approximations in the Baym-Kadanoff theory with the dynamical and conserving vertices in Sec. II. Qualitative differences between the critical behavior of the dynamical and conserving vertices are demonstrated on the elementary self-consistent approximate schemes in Sec. III.   We explicitly demonstrate the incompatibility of the stability criteria derived from the dynamical and conserving vertices on the SIAM in the strong-coupling Kondo limit within the fluctuation-exchange approximation in Sec. IV. Discussion and conclusions are presented in Sec. V. We added two appendices with supporting calculations and formulas. 

Although our explicit calculations were, for simplicity, done for a specific self-consistent dynamical approximation in the single-impurity model, the conclusions are general and hold for any approximation in the Baym-Kadanoff scheme, including the dynamical mean-field theory \cite{Georges:1996aa}, its ab initio extensions \cite{Lichtenstein:1998aa}, or its parquet extensions with the two-particle self-consistency \cite{Janis:1999aa,Rohringer:2018aa}.

%{\itshape Dynamical and conserving vertices.} 

\section{Baym-Kadanoff construction: Dynamical and conserving vertices}

The self-energy, the one-particle irreducible vertex, is standardly used as a generator of quantum many-body approximations. One needs, however, the corresponding thermodynamic functional connecting the heat effects with the external forces comprised in the self-energy.  It exists for the so-called $\Phi$-derivable approximations in the form of the  Luttinger-Ward functional $\Phi[U;G]$ with the bare interaction $U$ and the renormalized one-particle propagator $G$. Such approximations are thermodynamically consistent and obey macroscopic conservation laws with appropriately defined conserving vertices.  The generating thermodynamic functional of the Baym-Kadanoff scheme with the spin-dependent self-energy $\Sigma_{\sigma}$ and the one-particle Green function $G_{\sigma}$ can be represented on a regular lattice as follows    
\begin{multline}\label{eq:PhiG}
 \frac 1{{ N}}W[\Sigma ,G]  =\Phi [U;G]  - \sum_{\sigma = \pm 1}\frac 1{\beta}\sum_{n=-\infty}^{\infty}\frac 1{N}\sum_{\bf k}
 \\
   \left\{      
 e^{i\omega_n0^{+}}\ln \left[ i\omega _n+\mu _\sigma -\epsilon  
     ({\bf k}) -\Sigma_\sigma ({\bf
       k},i\omega_n)\right] 
          \right. \\  \left.      \phantom{ e^{i\omega_n0^{+}}}    
       +\ {G}_\sigma ({\bf k}, i\omega_n)\Sigma _\sigma ({\bf k},i\omega_n) \right\}  \,,
\end{multline}
where $N$ is the number of lattice sites, $\beta = 1/k_{B}T$, $\omega_{n} = (2n + 1)\pi k_{B}T$ are fermionic Matsubara frequencies, and $\epsilon(\veck)$  is the lattice dispersion relation.  We denoted $\mu_{\sigma} = \mu + \sigma h$, with $\mu$ the chemical potential and $h$ the external Zeeman magnetic field. We further resort to single-orbital models with a local Hubbard interaction $U$ in the tight-binding description. 

\begin{figure} 
\hspace*{-10pt}\includegraphics[width=9cm]{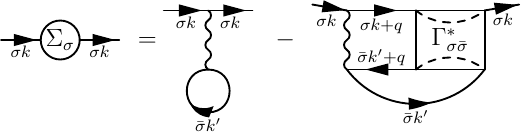} 
\caption{Diagrammatic representation of the Schwinger-Dyson equation,~\eqref{eq:SDE-symbolic} with the horizontal electron-hole propagation $\star$ with a conserving transfer bosonic momentum $\vecq$ and frequency $\nu_{m} = 2m \pi k_{B}T$. Arrows indicate the charge propagation; the electron propagates from left to right while the hole from right to left. We abbreviated independent integration variables in this integral equation  $k=(\veck,i\omega_{n})$, $q=(\vecq,i\nu_{m})$.   \label{fig:SDE}}
\end{figure}

The equilibrium states are determined from thermodynamic functional $W[\Sigma ,G]$ via stationarity with respect to variations of complex variables $\Sigma_\sigma ({\veck},i\omega_n)$ and $G_\sigma ({\veck},i\omega_n)$. The first stationarity condition $\delta W[\Sigma ,G]/\delta{\Sigma}(\veck,\omega_{n}) = 0$ leads  to the Dyson equation $G_\sigma ({\veck},i\omega_n) = [i\omega _{n} + \mu_{\sigma} - \epsilon(\veck) - \Sigma_\sigma ({\veck},i\omega_n)]^{-1}$. The second stationarity equation  determines the self-energy that can be represented in the form of the Schwinger-Dyson equation
\begin{multline}\label{eq:SDE-symbolic}
\Sigma_{\sigma}[U;G] = \frac{\delta \Phi[U;G]}{\delta {G}_{\sigma}} 
=  U\left\langle {G}_{\bar{\sigma}}\right\rangle
\\
 -\ U G_{\sigma}{G}_{\bar{\sigma}}\star\Gamma^{\ast}_{\sigma\bar{\sigma}}[U;G] \circ G_{\bar{\sigma}} \,. 
\end{multline}
The angular brackets denote the normalized sum over the fermionic Matsubara frequencies and momenta, $ \langle X\rangle = (\beta N)^{-1}\sum_{\omega_{n},\veck}X(\veck,i\omega_{n})$. Symbols $\star$ and $\circ$ represent two-particle and one-particle convolutions in a specific scattering channel propagating the singlet electron-hole pairs with $\bar{\sigma} = - \sigma$, respectively. The Schwinger-Dyson equation is represented diagrammatically in  Fig.~\ref{fig:SDE}.  It separates the two-particle contribution from the one-particle one to the self-energy. The two-particle contribution is convolution of the electron-hole propagator  $G_{\sigma}G_{\bar{\sigma}}\star = (\beta N)^{-1}\sum_{\veck,\omega_{n}}G_{\sigma}(\veck,i\omega_{n})G_{\bar{\sigma}}(\veck + \vecq,i\omega_{n} + i\nu_{m})$  and the one-particle irreducible singlet electron-hole vertex  $\Gamma^{*}$.  

Vertex $\Gamma^{*}$ can further be decomposed by introducing a two-particle vertex $\Lambda^{\ast}$ irreducible with respect to the single electron-hole propagation. We then obtain a Bethe-Salpeter equation in the singlet electron-hole channel
\begin{multline}\label{eq:BSE-eh-symbolic}
\Gamma^{\ast}_{\sigma\bar{\sigma}}[U;G] = \Lambda^{\ast}_{\sigma\bar{\sigma}}[U;G]
\\
  - \Lambda^{\ast}_{\sigma\bar{\sigma}}[U;G]G_{\sigma}{G}_{\bar{\sigma}}\star\Gamma^{\ast}_{\sigma\bar{\sigma}}[U;G]  \,.
\end{multline}
Vertex $\Lambda^{\ast}_{\sigma\bar{\sigma}}$ is a sum of all diagrams that cannot be disconnected by cutting simultaneously electron and hole lines; it is the electron-hole irreducible vertex. In this respect, this vertex is fully diagrammatically controlled and is used to generate dynamical approximations in the Baym-Kadanoff self-consistent scheme.  

 It is important to mention that vertex $\Gamma^{\ast}$ is not directly derived from the generating thermodynamic functional of Eq.~\eqref{eq:PhiG}. It was derived from a specific representation of the self-energy in the form of the Schwinger-Dyson equation. 
 It, hence, does not guarantee conservation laws and sum rules. The irreducible vertex $\Lambda_{\sigma\bar{\sigma}}[U;G]$ should be connected in the conserving theory with  the self-energy via a functional Ward identity\cite{Baym:1962aa}
\begin{equation}\label{eq:WI-functional}
\Lambda_{\sigma\bar{\sigma}}^{W}[U;G] =  \frac{\delta \Sigma_{\sigma}[U;G]}{\delta {G}_{\bar{\sigma}}} \,. 
\end{equation}
It is the two-particle irreducible vertex derived as a second variational derivative from the Luttinger-Ward functional. When used in the corresponding Bethe-Salpeter equation, we obtain the conserving vertex $\Gamma_{\sigma\bar{\sigma}}[U;G]$. The dynamical vertex $\Gamma_{\sigma\bar{\sigma}}^{\ast}$  from the Schwinger-Dyson and Bethe-Salpeter equations, Eqs.~\eqref{eq:SDE-symbolic} and~\eqref{eq:BSE-eh-symbolic} should equal the conserving one, $\Gamma_{\sigma\bar{\sigma}}$ from the Bethe-Salpeter equation with the conserving irreducible vertex $\Lambda^{W}_{\sigma\bar{\sigma}}$ in the exact theory. That is $\Gamma^{\ast}[U;G]  = \Gamma[U;G]$. This cannot, however, be achieved in accessible approximate treatments.  

Consistency between the one and two-particle functions in the Baym-Kadanoff construction with the generating self-energy functional  $\Sigma[G]$  is guaranteed if the functional Ward identity, Eq.~\eqref{eq:WI-functional}, is obeyed and the full vertex is represented via the Bethe-Salpeter equation~\eqref{eq:BSE-eh-symbolic}. On the other hand, the Schwinger-Dyson equation determines the self-energy being the sum of all Feynman diagrams and is the complete diagrammatic solution of the many-body perturbation theory. To guarantee equality $\Gamma^{\ast}[U;G]  = \Gamma[U;G]$, the two-particle irreducible vertex $\Lambda_{\sigma\bar{\sigma}}$ must comply with another integral equation with its functional derivative  when Eq.~\eqref{eq:BSE-eh-symbolic} is inserted in Eq.~\eqref{eq:SDE-symbolic}
\begin{multline}\label{eq:SDE-derivative}
 \Lambda^{W}_{\sigma\bar{\sigma}} = \Lambda^{\ast}_{\sigma\bar{\sigma}} = U - U\left[ 1 + G_{\sigma}{G}_{\bar{\sigma}}\Lambda^{\ast}_{\sigma\bar{\sigma}}\star\right]^{-1}G_{\sigma}\left\{\Lambda^{\ast}_{\sigma\bar{\sigma}}\phantom{\frac 12}
\right. \\ \left.
 +\ {G}_{\bar{\sigma}}\frac{\delta\Lambda^{\ast}_{\sigma\bar{\sigma}}}{\delta {G}_{\bar{\sigma}}}\right\}
%\right. \\ \left.
\left[ 1 + \star G_{\sigma}{G}_{\bar{\sigma}}\Lambda^{\ast}_{\sigma\bar{\sigma}}\right]^{-1}\circ G_{\bar{\sigma}} \,.
%\\
%= \Lambda_{\sigma-\sigma}^{eh}[U;G,\overline{G}]\,.
\end{multline}

It is evident that Eqs.~\eqref{eq:WI-functional} and~\eqref{eq:SDE-derivative} cannot be obeyed simultaneously with approximate irreducible functions and $\Gamma^{\ast}[U;G]  \neq \Gamma[U;G]$.  To keep the approximation for the self-energy from Eq.~\eqref{eq:SDE-symbolic} conserving, one has to treat vertex $\Gamma^{\ast}$  as an auxiliary function and to give the physical meaning only to vertex $\Gamma$ from the Bethe-Salpeter equation~\eqref{eq:BSE-eh-symbolic}. This is, however, possible only up to a critical point, the divergence in the auxiliary vertex $\Gamma^{\ast}$. It happens if 
\be\label{eq:Lambda-Min}
\mathrm{Min}\ \mathrm{Sp}\left[\Lambda^{\ast} GG\star\right](\vecQ,0) = -1
\ee
for a given momentum $\vecQ$ determining the symmetry to be broken beyond this critical point. One cannot continue the approximation beyond the critical point unless the Ward identity is obeyed. At least to the extent that would guarantee that the critical point in the vertex function $\Gamma^{\ast}$ introduces a symmetry breaking and the continuous emergence of an order parameter in the self-energy.   The problem is, however, that the two vertex functions $\Gamma$ and $\Gamma^{\ast}$ in the $\Phi$-derivable approximate theories lead to two different critical points that should coincide in the exact solution. This conclusion does not depend on the way we reached the electron-hole irreducible vertex $\Lambda^{\ast}_{\sigma\bar{\sigma}}$ since its functional derivative in Eq.~\eqref{eq:SDE-derivative} generates new Feynman diagrams not contained in it \cite{Janis:1998aa}.
Equality $\Lambda^{\ast} = \Lambda^{W}$, that is, obeying  Eq.~\eqref{eq:SDE-derivative}, is beyond the reach of the diagrammatically controlled approximate schemes. The only feasible possibility is to select a starting irreducible vertex $\Lambda^{(0)}$ and iterate the consistency equation~\eqref{eq:SDE-derivative} successively. The difference between the critical behavior of vertices $\Gamma$ and $\Gamma^{\ast}$ is not removed in this iterative scheme.

%Critical behavior $\mathrm{min}\left[\Lambda GG\star\right] +1 = 0$
 
 \section{Conserving vertex in approximate schemes} 
 
 We demonstrate explicitly the difference between the dynamical and conserving vertices on simple approximations. We start with the simplest, static Hartree-Fock approximation. Its Luttinger-Ward functional is
 \begin{multline}
 \Phi_{HF}[U;G] = U\sum_{\sigma}\frac 1{\beta N}\sum_{\veck}\sum_{\omega_{n}}e^{i\omega_{n}0^{+}}G_{\sigma}(\veck,i\omega_{n})
 \\
 \times \frac 1{\beta N}\sum_{\vecp}\sum_{\omega_{l}}e^{i\omega_{l}0^{+}}{G}_{\bar{\sigma}}(\vecp,i\omega_{l}) \,.
 \end{multline}
 The self-energy is a number $\Sigma_{\sigma} = Un_{\bar{\sigma}}$. The dynamical vertex $\Gamma^{\ast} = 0$ is easily deduced from the corresponding Schwinger-Dyson equation. The conserving vertex, on the other hand, is non-trivial with $\Lambda^{W} =U$ as easily found from Eq.~\eqref{eq:WI-functional}. That is why the Hartree-Fock static mean-field approximation leads to a thermodynamic mean-field critical behavior. It contains no quantum dynamics.
 
 The Hartree-Fock approximation is a weak-coupling static solution missing the quantum fluctuations due to the non-commutativity of the Hamiltonian of the kinetic energy and the interaction Hamiltonian. The next step beyond the static approximation is fluctuation exchange (FLEX) \cite{Bickers:1989aa,Bickers:1989ab,Bickers:1991aa,Janis:1998aa}, where we use the bare interaction as the two-particle irreducible vertex in the Schwinger-Dyson equation,  $\Lambda^{\ast} =U$. The corresponding Luttinger-Ward functional reads
 \begin{multline}
 \Phi_{FLEX}[U;G] 
 \\
 = \frac 12\sum_{\sigma}\frac 1{\beta N}\sum_{\vecq}\sum_{\nu_{m}}
\ln\left[1 + U\phi_{\sigma\bar{\sigma}}(\vecq,i\nu_{m}) \right] \,,
 \end{multline}
where we introduced a singlet electron-hole bubble 
 \begin{multline}
\phi_{\sigma\bar{\sigma}}(\vecq,i\nu_{m}) 
\\
= \frac 1{\beta N} \sum_{\veck}\sum_{\omega_{n}}G_{\sigma}(\veck,i\omega_{n}){G}_{\bar{\sigma}}(\veck + \vecq,i\omega_{n} + i \nu_{m}) \,.
 \end{multline}
Notice the symmetry of this bubble $\phi_{\bar{\sigma}\sigma}(\vecq,i\nu_{m}) = \phi_{\sigma\bar{\sigma}}(-\vecq,-i\nu_{m})$.

The Schwinger-Dyson equation for the self-energy 
 \begin{multline}
\Sigma_{\sigma}(\veck,i\omega_{n})  =  
\frac U{\beta N}\sum_{\vecp}\sum_{\omega_{l}}e^{i\omega_{l}0^{+}}{G}_{\bar{\sigma}}(\vecp,i\omega_{l})
\\
- \frac {U^{2}}{\beta N} \sum_{\vecq}\sum_{\nu_{m}}\frac{\phi_{\sigma\bar{\sigma}}(\vecq,i\nu_{m}){G}_{\bar{\sigma}}(\veck + \vecq,i\omega_{n} + i \nu_{m})}{1 + U\phi_{\sigma\bar{\sigma}}(\vecq,i\nu_{m})}
 \end{multline}
contains a non-trivial dynamical vertex $\Gamma^{\ast}$ that may become divergent. According to Eq.~\eqref{eq:Lambda-Min}, it happens if $U\phi(\vecQ,0) = -1$ for an appropriate vector $\vecQ$ determining the type of the magnetic instability.   

The real magnetic instability is, however, determined from the divergence of the magnetic susceptibility, $\chi = dm/d h$ evaluated at $h=0.$ The magnetic susceptibility depends then on the derivative of the self-energy as $\chi = - \sum_{\sigma} \sigma \langle G^{2}d\Sigma_{\sigma}/ d h\rangle$. To evaluate this derivative, we need to evaluate the conserving irreducible vertex $\Lambda^{W}$ in the FLEX approximation. It is
 \begin{multline}\label{eq:LambdaW-FLEX}
\Lambda^{W}_{\sigma\bar{\sigma}}(\veck,i\omega_{n}, \veck^{\prime},i\omega_{n^{\prime}};\vecq,i\nu_{m}) 
 =  \frac{\delta \Sigma_{\sigma}(\veck,i\omega_{n})}{\delta {G}_{\bar{\sigma}}(\veck + \vecq,i\omega_{n} + i\nu_{m})}
 \\
  = \frac U{1 + U\phi_{\sigma\bar{\sigma}}(\vecq,i\nu_{m})} 
-\ \frac{U^{2}}{\beta N}\sum_{\vecQ}\sum_{\nu_{l}} 
\\
\frac{G_{\sigma}(\veck + \vecq -  \vecQ,i\omega_{n} + i\nu_{m} - i\nu_{l}){G}_{\bar{\sigma}}(\veck + \vecQ,\i\omega_{n} + i\nu_{l})}{\left[1 + U\phi_{\sigma\bar{\sigma}}(\vecQ,i\nu_{l})\right]^{2}} \,.
 \end{multline}
 The critical point for this irreducible vertex is found as the minimum eigenvalue of a matrix $\Lambda^{W}_{\sigma\bar{\sigma}}(\veck,i\omega_{n}, \veck^{\prime},i\omega_{n^{\prime}};\vecQ,0)G_{\sigma}(\veck^{\prime},i\omega_{n^{\prime}})G_{\bar{\sigma}}(\veck^{\prime} + \vecQ,i\omega_{n^{\prime}})$  in variables $\veck, i\omega_{n}$ and $\veck^{\prime},i\omega_{n^{\prime}}$  for a specific vector $\vecQ$.  To determine the susceptibility, we need to evaluate also the triplet irreducible vertex $\delta \Sigma_{\sigma}/\delta G_{\sigma}$. It compensates the second term on the right-hand side of the in Eq.~\eqref{eq:LambdaW-FLEX} in the susceptibility of the spin-symmetric solution.  
 
 The difference between the dynamical and conserving irreducible vertices is qualitative. The dynamical irreducible vertex is a number, the bare interaction $U$, while the conserving irreducible vertex $\Lambda^{W}$ is a potentially divergent function in the strong-coupling regime. It is indeed the case, as we demonstrate in a generic example of the SIAM in the strong-coupling Kondo regime.

 \section{Single impurity Anderson model: Kondo scales}  
 
The strong-coupling regime of the single impurity Anderson model at half-filling is the simplest situation for testing the consistency of approximate solutions of critical behavior. The exact spin-symmetric solution is stable for interaction strengths with an exponentially small Kondo scale. Consistent approximations should deliver the same critical Kondo scale in the magnetic susceptibility, spectral function, and specific heat when approaching zero temperature and infinite interaction strength. Each thermodynamic quantity can then be used for finding the stability criterion of the high-temperature, spin-symmetric solution. The static Hartree approximation predicts instability of the spin-symmetric solution in the strong-coupling regime. On the other hand, FLEX approximations lead to a stable spin-symmetric solution at any interaction strength when deduced from the dynamical vertex. This conclusion is not, however, confirmed by the conserving vertex.   
 
 \subsection{Self-energy in the FLEX approximation}
 
The SIAM contains only local quantum dynamics with Matsubara frequencies as the only dynamical variables. The dispersion relation is replaced in this model by energy $\Delta$, forming a gap around the real axis for the Matsubara frequencies. The full Green function of the SIAM is 
 \begin{subequations}
\begin{multline}
G_{\sigma}(i\omega_{n}) 
\\
= \frac1{i\left(\omega_{n}  + \mathrm{sign}(\omega_{n})\Delta\right)  + \bar{\mu}  + \sigma\bar{h} - \Sigma_{\sigma}(i\omega_{n})} \,,
\end{multline}
where we used the solution of the Hartree approximation with charge and spin densities $n$ and $m$, respectively. We denoted $\bar{h} = h +  U m/2$ and $\bar{\mu} = \mu - Un/2$.

The Schwinger-Dyson equation analytically continued to real frequencies for the SIAM is 
\begin{multline}\label{eq:Sigma-FLEX}
\Sigma_{\sigma}(\omega) 
%= - \frac{U^{2}}{2\beta} \sum_{\nu_{m}} \frac{\phi(i\nu_{m})}{1 + U\phi(i\nu_{m}) } G_{\bar{\sigma}}(i\omega_{n} + i\nu_{m}) 
=  - U^{2}\int_{-\infty}^{\infty}\frac{dx}{\pi}\left\{ b(x) \Im\left[\frac{\phi(x)}{1 + U\phi(x) }\right]
\right. \\ \left. 
\times G_{\bar{\sigma}}(\omega + x)   - f(\omega + x)\frac{\phi^{*}(x)\Im G_{\bar{\sigma}}(\omega + x)}{1 + U\phi^{*}(x)}
\right\} \,,
\end{multline}
\end{subequations}
where the electron-hole bubble has the following representation
\begin{multline}
\phi(\omega)  = \frac 12\sum_{\sigma}\int_{-\infty}^{\infty}\frac{d\omega}{\pi}f(\omega)
\\
\times \left[G_{\sigma}(\omega + x) 
+ G_{\sigma}(\omega - x)\right]\Im G_{\bar{\sigma}}(\omega) \,.
\end{multline}
The frequency variables in the above and all following expressions are taken as the limit from the upper complex half-plane, $\omega= \omega + i0^{+}$, $x = x + i0^{+}$,  and $\phi^{*}$ is complex conjugate. 

This approximation is equivalent to the renormalized random phase approximation (RRPA)  or the ladder approximation introduced by Suhl \cite{Suhl:1967aa}.  It was used to analyze the local magnetic moments' behavior and to calculate the magnetic susceptibility \cite{Levine:1968aa}.  Due to the one-particle self-consistency, the dynamical vertex of the RRPA, singlet ladder approximations or FLEX  suppresses the Hartree instability. Hamann derived their strong-coupling asymptotics \cite{Hamann:1969aa}. The Kondo scale calculated from the dynamical vertex appeared to be $\ln a = -(U\rho_{0})^{2}/3$. It differs from the exact result but shares with it the same critical point at $T=0$ and $U_{c}=\infty$. We disclosed that the difference in the critical asymptotics is due to the lack of a two-particle self-consistency \cite{Janis:1999aa}. We restored the correct linear dependence of the logarithm of the  Kondo scale on the interaction strength via reduced parquet equations for the two-particle-irreducible vertex in the electron-hole scattering channel \cite{Janis:2007aa,Janis:2008ab,Janis:2017aa,Janis:2017ab,Janis:2019aa}. 
%Based on these results, it was concluded that the spin-symmetric state of the SIAM in the renormalized conserving approximations remains stable in the whole range of the interaction strength. The eventual conclusion about the stability of equilibrium states cannot, however, be made without checking on the conserving vertex.

 The improper usage of the dynamical vertex, which does not obey the Ward identity, to determine thermodynamic stability was pointed out in the literature \cite{Haussmann:2007aa,He:2016aa}. It was recognized early that the magnetic susceptibility contains additional diagrams not included in the dynamical vertex \cite{Levine:1968aa}. It is, however, tricky to derive the full expression for the susceptibilities calculated from the conserving vertex where uncontrolled Feynman diagrams from the functional derivative of the self-energy must be added to the dynamical vertex \cite{Janis:1998aa}.  That is why the dynamical vertex defined via a controlled sum of explicitly selected Feynman diagrams has been used to evaluate both the self-energy and the response functions in the self-consistent approximations \cite{Bickers:1989ab,Georges:1996aa,Moriya:2000aa,Kemper:2010aa,Onari:2012aa,Hirschmeier:2015aa,Re:2021aa,Witt:2021aa,Klett:2022aa,Witt:2023aa}. The situation does not improve even if we extend the Luttinger-Ward functional by two-particle vertex functions via the De Dominicis-Martin scheme leading to a parquet construction \cite{DeDominicis:1964aa,DeDominicis:1964ab,Janis:1999aa,Li:2016aa,Rohringer:2018aa,Li:2020aa}.

\subsection{Quantum criticality - Polar approximation}

The FLEX or RRPA of the SIAM allows us to evaluate the conserving vertex, at least in the strong-coupling Kondo limit, explicitly. We use the asymptotic form of the solution for the spin-polarized self-energy of the SIAM in the strong-coupling limit as done in Refs.~\cite{Hamann:1969aa,Janis:2007aa}. Since the divergence of the dynamical vertex leads to a non-integrable singularity in the Schwinger-Dyson equation, the exact strong-coupling limit can be obtained from the leading low-frequency asymptotics of the dynamical vertex, the so-called polar approximation \cite{Janis:2007aa}. To check the stability of the spin-symmetric solution of FLEX in the strong-coupling regime and to obtain the magnetic susceptibility, we use the polar approximation, asymptotically exact in the Kondo regime. It is characterized by a magnetic Kondo scale near the magnetic instability. Only if the Kondo scales from the dynamical and conserving vertices behave qualitatively in the same way the thermodynamic critical behavior is consistent with quantum dynamics. 

We set $T=0$ and take the limit $U\to\infty$ of the half-filled SIAM to reach the Kondo critical behavior. The pole of the dynamical vertex of the RRPA emerges when the Kondo scale in the denominator of the dynamical vertex at zero transfer frequency  $a = 1 + U\phi(0) \to 0$, where $\phi(\omega)$ is the singlet electron-hole bubble, $\phi(0) = - \int_{-\infty}^{0}d\omega \Im\left[G_{\uparrow}(\omega)G_{\downarrow}(\omega)\right]$. Since the integral in the Schwinger-Dyson equation is logarithmically infrared divergent, only low frequencies in the dynamical vertex deliver the dominant contribution to the self-energy.  The one-particle Green function is analytic in complex frequencies $\omega$, and we can set the integration variable of the Schwinger-Dyson equation $x=0$ in it. We thus turn the integral equation for the self-energy algebraic in the one-electron variables with separate two-particle integrals \cite{Janis:2007aa}
\begin{subequations}\label{eq:Sigma-Algebraic}
\begin{multline}\label{eq:ReSigma-Algebraic}
\Re\Sigma_{\sigma}(\omega) = -U^{2}\Re {G}_{\bar{\sigma}}(\omega) \int_{-\infty}^{\infty}\frac{dx}{\pi} b(x)\Im \left[\frac{\phi(x) }{1 + U\phi(x) }\right]
\\
+\ U^{2}\Im {G}_{\bar{\sigma}}(\omega) \int_{-\infty}^{\infty}\frac{dx}{\pi} f(\omega + x) \Re \left[\frac{\phi(x) }{1 + U\phi(x) }\right] \,,
\end{multline}
\begin{multline}\label{eq:ImSigma-Algebraic}
\Im\Sigma_{\sigma}(\omega) = -U^{2}\Im {G}_{\bar{\sigma}}(\omega) 
\\
\times\int_{-\infty}^{\infty}\frac{dx}{\pi} \left(b(x) + f(\omega + x)\right)\Im \left[\frac{\phi(x) }{1 + U\phi(x) }\right] \,.
\end{multline}
\end{subequations}
These algebraic equations for the real and imaginary parts of the self-energy deliver the exact solution of the FLEX approximation in the leading order in the vanishing Kondo scale $a\to 0$.  We can approximate the electron-hole bubble in the dynamical vertex in this limit by its low-frequency asymptotics
\be
\phi(\omega) \doteq \phi(0) - i\phi^{\prime}\omega \,,
\ee
 with $\phi^{\prime}= \pi\rho_{0}^{2}$ independent of the interaction strength and $\rho_{0}$ the bare density of states. 

 Equations~\eqref{eq:Sigma-Algebraic} deliver the correct strong-coupling asymptotics of the dynamical self-energy. They can be extended outside this limit. Their solution obeys the Fermi-liquid properties at low temperatures, and the algebraic approximation can be used as a suitable interpolation between the weak and strong couplings of the SIAM.  To extend qualitatively correctly the algebraic approximation to the weak-coupling regime $U\to 0$, we need to introduce a high-frequency cutoff $\Omega$ in the two-particle integrals. The cutoff does not affect the diverging part of the integral and the critical behavior in the strong-coupling limit. 

The explicit integrations of the two-particle contribution to the self-energy are presented in Appendix A. The resulting algebraic equations for the real and imaginary parts of the self-energy read 
\begin{subequations}\label{eq:Sigma-phiprime}
\begin{align}\label{eq:ReSigma-phiprime}
\Re \Sigma_{\pm}(\omega) &=   \frac 1{\pi\phi^{\prime}}\left\{\ln\sqrt{1 + \frac{U^{2}\phi^{\prime 2}\Omega^{2}}{a^{2}}} \Re G_{\mp}(\omega) 
\right. \nonumber \\
&\left. +\ (1 - a)\arctan\left(\frac{U\phi^{\prime}\omega}a\right) \Im G_{\mp}(\omega)\right\}\,,
\\ \label{eq:ImSigma-phiprime}
\Im \Sigma_{\pm}(\omega) &= \frac 1{\pi\phi^{\prime}}\ln\sqrt{1 + \frac{U^{2}\phi^{\prime 2}\omega^{2}}{a^{2}}}\Im G_{\mp}(\omega)\,,
\end{align}
\end{subequations}
where subscript $\pm$ refers to the spin variable $\sigma=\pm$.

\subsection{Solution of the polar approximation}

We derive the solution of the polar approximation for arbitrary interaction strength. 
We introduce new variables for convenience and simplification of the expressions and set  $\Delta = 1$, $X(\omega) = - \Re\Sigma(\omega)$ and $Y(\omega)= - \Im\Sigma(\omega)$: 
\begin{subequations}\label{eq:STZ-Def}
\begin{align}
S(\omega) &= \omega + X(\omega)\,,
\\
T(\omega) &= 1 + Y(\omega)\,,
\\
Z(\omega)^{2} & = S(\omega)^{2} + T(\omega)^{2}\,,
\\
l_{\Omega} &= \ln\sqrt{1 + \bar{U}^{2}\Omega^{2}} \,,
\\
A(\omega) &= \left(1 - a\right)\arctan\left(\bar{U}\omega\right) \,,
\\
L(\omega) &=  \ln\sqrt{1 + \bar{U}^{2}\omega^{2}}\,,
\end{align}
\end{subequations}
where we used a rescaled interaction strength $\bar{U}= U/\pi a$.

The equations for functions $T(\omega)$ and $S(\omega)$ read
\begin{subequations}\label{eq:ST-Z}
\begin{align}
T(\omega)&= \frac{Z(\omega)^{2}}{Z(\omega)^{2} - L(\omega)}\,,
\\
S(\omega)&= T(\omega)\frac{\omega\left(Z(\omega)^{2} - L(\omega)\right) + A(\omega)}{Z(\omega)^{2} + l_{\Omega}}\,.
\end{align}
\end{subequations}
Using these two equations in Eqs.~\eqref{eq:STZ-Def}, we obtain an algebraic equation for the square norm of the inverse Green function $Z(\omega)^{2}$
%\begin{subequations}\label{eq:Z-norm}
%\be\label{eq:Z-norm1}
%\left(Z(\omega)^{2} - L(\omega)\right)^{2}\left(Z(\omega)^{2} + l_{\Omega}\right)^{2} = Z(\omega)^{2}\left\{\left(Z(\omega)^{2} + l_{\Omega}\right)^{2} + \left[\omega\left(Z(\omega)^{2} - L(\omega)\right) + A(\omega)\right]^{2}\right\}\,.
%\ee
%Alternatively, we can rewrite into a more suitable form
\begin{multline}\label{eq:Z-norm2}
Z(\omega)^{2}  = \frac{1}{\left(1 - L(\omega)/Z(\omega)^{2}\right)^{2}}\left\{1 \phantom{\frac12}
\right. \\ \left.
+\ \omega^{2}\frac{\left[1 - \left(L(\omega) - \bar{A}(\omega)\right)/Z(\omega)^{2}\right]^{2}}{\left(1 + l_{\Omega}/Z(\omega)^{2}\right)^{2}}\right\} \,,
\end{multline}
%\end{subequations}
where we denoted $\bar{A}(\omega) = A(\omega)/\omega$.
 
 %We solve this equation by halving the interval between the upper and lower bounds for $Z(\omega)^{2}$ of the right-hand side. 
 We use $\bar{U}= U/\pi a$ and $1/a$ as independent variables. The interaction enters Eqs.~\eqref{eq:STZ-Def} only as $\bar{U}$ and the resulting equation for $1/a$ then is
\begin{multline}\label{eq:barUaInverse}
\frac 1a = 1 
\\
-\  2\bar {U}\int_{-\infty}^{0}d\omega \frac{\omega\left(Z(\omega)^{2} - L(\omega)\right) + \bar{A}(\omega)}{\left(Z(\omega)^{2} - L(\omega)\right)^{2}\left(Z(\omega)^{2} + l_{\Omega}\right)} \,.
\end{multline}
%\end{subequations}  
Consequently, variable $\omega$ enters functions $Z(\omega)$, $L(\omega)$, and $A(\omega)$ only as powers of its square, that is, $\omega^{2}$ becomes the elementary variable. 
%Switching to a new variable $x=\omega^{2}$, we obtain
%\be\label{eq:barUaInverse}
%\frac 1a = 1 -  \bar {U}\int_{0}^{\infty}dx \frac{\left(\bar{Z}(x)^{2} - \bar{L}(x)\right) + \bar{A}(x)}{\left(\bar{Z }(x)^{2} - \bar{L}(x)\right)^{2}\left(\bar{Z}(x)^{2} + l_{\Omega}\right)}
%\ee
% where the bar denotes the dependence on $x=\omega^{2}$ of the the original functions. 

 It can be easily demonstrated that the polar approximation leads to Fermi liquid in weak coupling. The algebraic equations for the self-energy can be solved numerically for any interaction strength. Before we do so, we first assess its weak and strong coupling limits that can be solved analytically.  We use the weak-coupling solution to fix the integration cutoff $\Omega=\Delta$. We choose it to fit the weak-coupling value of the derivative $d\Sigma(\omega)/d\omega$ at $\omega=0$ to best reproduce the Fermi-liquid regime. The weak-coupling asymptotics is obtained from the limit of the Kondo scale $a\to 1$. 
 
 Fundamental equation~\eqref{eq:Z-norm2} for the square of the inverse Green function linearizes in the weak coupling to 
\be \label{eq:Z-Weak}
Z(\omega)^{2} = \omega^{2} + 1 + 2\frac{L(\omega) - \left(l_{\Omega} - \bar{A}(\omega)\right)\omega^{2}}{\omega^{2} +1}\,. 
\ee
Its solution in the leading order $a\to 0$ has the following input on its right-hand side 
\begin{subequations}
\begin{align}
L(\omega) &= \frac{U^{2}\omega^{2}}{2\pi^{2}}\,,
\\
l_{\Omega} &= \frac{U^{2}\Omega^{2}}{2\pi^{2}}\,,
\\
\bar{A}(\omega) &= \frac{U^{2}}{\pi^{2}}\,.
\end{align}
\end{subequations} 
%The solution of Eq.~\eqref{eq:Z-Weak} then is
%
The real and imaginary parts of the inverse Green function in this limit are
\begin{subequations}
\begin{align}
S(\omega)  &= \omega\left[1 - \frac{ l_{\Omega} - \bar{A}(\omega)}{\omega^{2} + 1}\right]\,,
\\
T(\omega) &= 1 + \frac{L(\omega)}{\omega^{2} + 1} \,.
\end{align}
\end{subequations}

Evaluating the derivative $d\Sigma(\omega)/d\omega$ in this approximation and comparing it with second-order perturbation theory we obtain, see Appendix B1, 
\begin{align}
\left.\frac{d X(\omega)}{d\omega}\right|_{\omega=0} &= \frac{U^{2}}{\pi^{2}}\left(1 - \frac{\Omega^{2}}2\right) \approx 0.533  \frac{U^{2}}{\pi^{2}}\,,
\end{align} 
from which the cutoff is
\begin{align}
\Omega &\approx 0.967\,.
\end{align} 

The strong-coupling solution corresponds to the limit $a\to 0$ and $1/a\gg \Omega$. It is the limit where the polar approximation reaches asymptotically the full solution of the convolutive Schwinger-Dyson equation. The equations for functions $X(\omega)$ and $Y(\omega)$ simplify in the strong-coupling limit and in the linear order of the magnetic field to
%\begin{subequations}
\begin{equation}
X_{+}(\omega) = -  \left[|\ln a|\Re G_{-}(\omega) 
%\right. \\ \left. 
+ \frac{\pi}2\mathrm{sign}(\omega)\Im G_{-}(\omega)\right]\,, 
\end{equation}
\begin{align}
Y(\omega) &= - \ln\sqrt{1 + \frac{U^{2}\omega^{2}}{a^{2}}}\Im G(\omega)\,.
\end{align}
%\end{subequations} 
The dependence on the cut-off frequency $\Omega$ vanishes, as expected.  The solution for the Kondo scale $a\to 0$ obeys the following equation in the strong-coupling limit 
\be\label{eq:a-strong-equation}
1 = \frac{4U}{\pi\sqrt{|\ln a|}}\int_{0}^{1}dx x\sqrt{1 - x^{2}} =  \frac{4U}{3\pi\sqrt{|\ln a|}} \,.
\ee
Its solution is 
\be\label{eq:a-asymptotic}
a = \exp\left\{- \frac{16U^{2}}{9\pi^{2}}\right\}\,.
\ee
See Appendix B2 for the detailed derivation. The power of the interaction strength in the logarithm of the Kondo scale agrees with Hamann's result \cite{Hamann:1969aa}.

\begin{figure*}
\includegraphics[width=17cm]{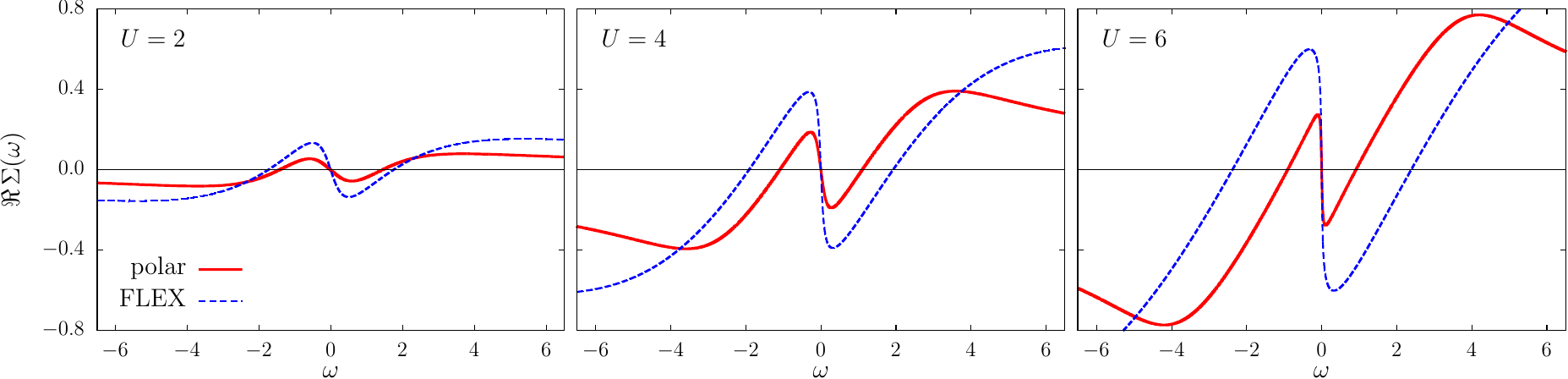}
\caption{Real part of the self-energy calculated for three interaction strengths, $U=2,4,6$ with $\Delta=\Omega=1$. We compared two solutions, the full convolutive equation, Eq.~\eqref{eq:Sigma-FLEX}  (dashed curves) and the self-energy from the polar approximation, Eqs.~\eqref{eq:Sigma-phiprime}  (solid curves).    \label{fig:RESigma}}
\end{figure*}

\begin{figure*}
\includegraphics[width=17cm]{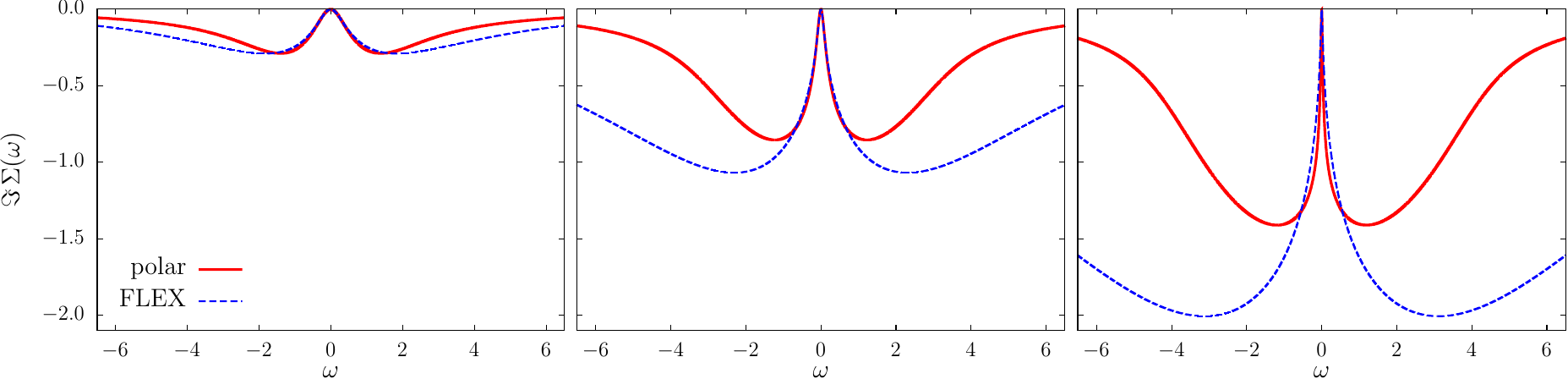}
\caption{Imaginary part of the self-energy for the full solution and the polar approximation. The setting is the same as in Fig.~\ref{fig:RESigma}.    \label{fig:ImSigma}}
\end{figure*}

 The numerical solution of Eqs.~\eqref{eq:Sigma-phiprime} was obtained on a standard PC via a Python code utilizing the \texttt{scipy} library. Eq.~\eqref{eq:Z-norm2} was solved via a standard root finding technique using the Krylov approximation for inverse Jacobian as implemented in the \texttt{scipy.optimize.root} function. Eq.~\eqref{eq:barUaInverse} was then solved using a simple iteration technique. We tested the polar approximation for several interaction strength values and compared the resulting self-energy with the full convolutive solution. We plotted the two self-energies in Figs.~\ref{fig:RESigma} and~\ref{fig:ImSigma} for $U=2,4,6$ standing for weak, intermediate, and strong coupling. We can see that the polar approximation fits the convolutive solution quite accurately for low frequencies near the Fermi energy for all interaction strengths. Notice that the polar self-energy grows from zero slower in weak coupling and faster in strong coupling. The polar approximation contains fewer fluctuations in weak coupling than the convolutive one and approaches strong coupling faster than the convolutive solution.   

\begin{figure}
\includegraphics[width=85mm]{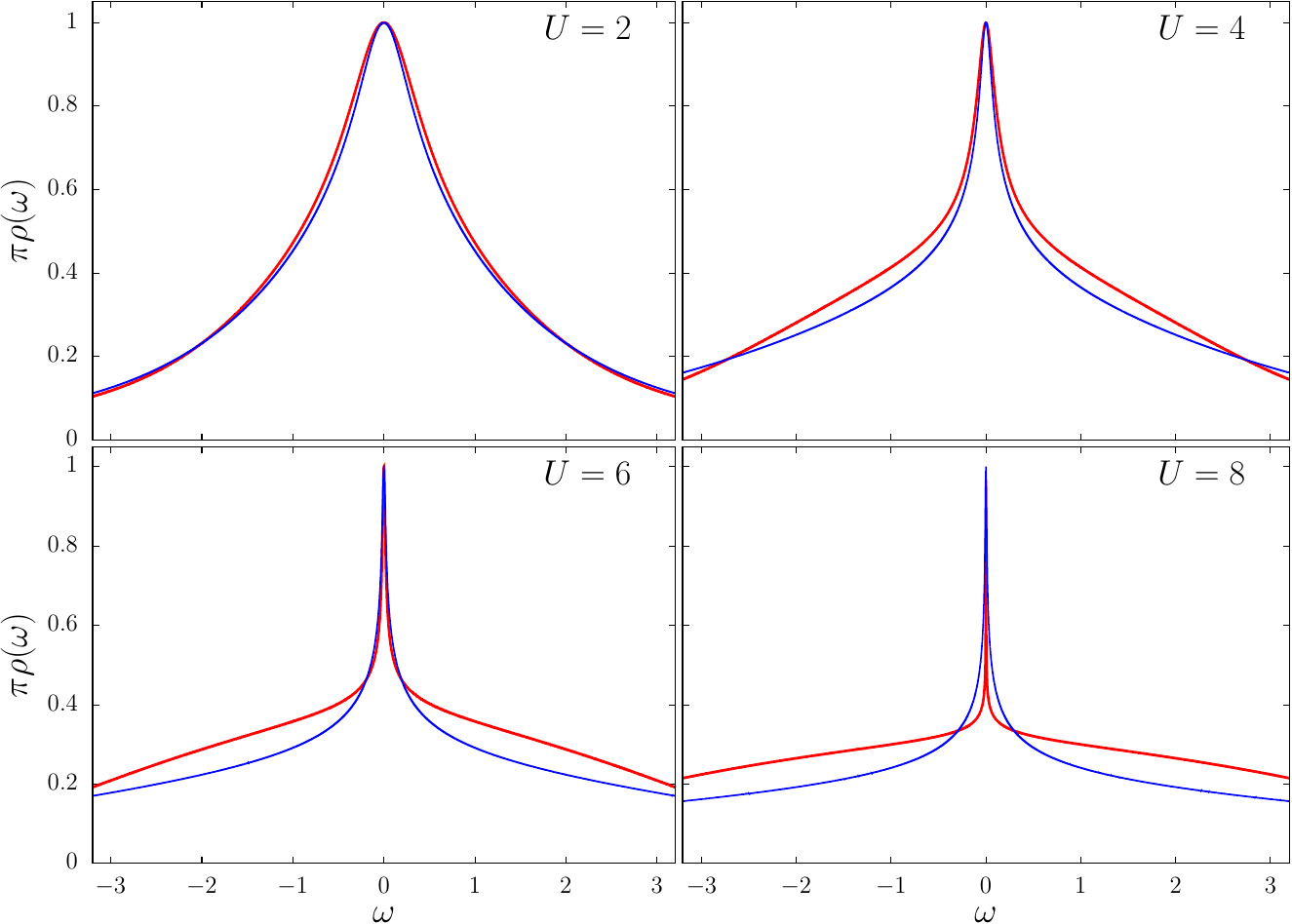}
\caption{Spectral function at zero temperature calculated with the FLEX self-energy, blue curves, compared with the approximate one from Eq.~\eqref{eq:Sigma-phiprime}, red curves, in weak and intermediate coupling. The cutoff frequency was chosen $\Omega=\Delta=1$.    \label{fig:DOS}}
\end{figure}

We can conclude that the algebraic approximation reproduces the solution of the full integral Schwinger-Dyson equation for the self-energy quite well in the whole range of the interaction strength. The agreement with the convolutive solution in the spectral function is acceptable, in particular, close to the Fermi energy, see Fig.~\ref{fig:DOS}. The two solutions slightly differ for higher frequencies $\omega \gtrapprox \Omega$ where neither displays the satellite Hubbard bands.  The algebraic solution approaches the strong coupling regime faster than the full convolutive one since it overestimates the impact of the diverging dynamical fluctuations from the critical region, where it becomes asymptotically exact.    

\begin{figure}
\includegraphics[width=8cm]{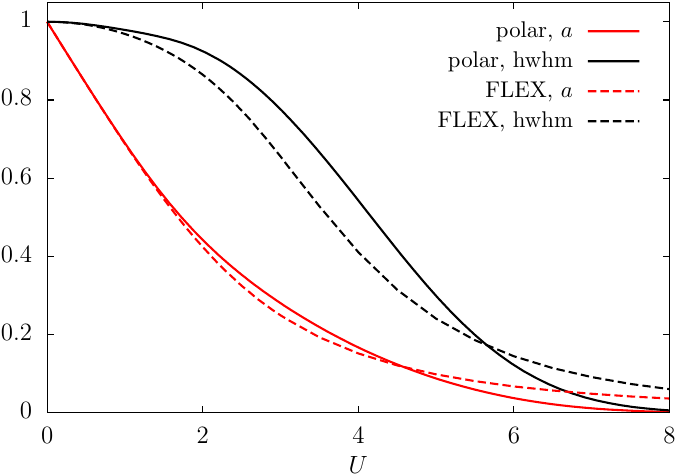}
\caption{The zero-temperature dimensionless Kondom scale $a= 1 + U\phi(0)$ compared with the half-width at half-maximum of the quasiparticle peak (hwhm) in the full convolutive solution (FLEX, dashed curves) and the polar approximation (solid curves).     \label{fig:UDep_a-hwhm}}
\end{figure}

We also compared the Kondo scale from the dynamical vertex $a=1 + U\phi(0)$ with the half-width of the quasiparticle peak taken at half-maximum of the spectral function (hwhm) calculated in the polar approximation and the convolutive solution (FLEX) with $\Delta=\Omega=1$ in Fig.~\ref{fig:UDep_a-hwhm}. The half-width of the quasiparticle peak is larger in the polar approximation than in the full convolutive solution for $U\lesssim5.65$ and smaller for larger values, which can also be seen in Fig.~\ref{fig:DOS}. The two scales differ in weak coupling, far from the Kondo regime, but approach each other in the strong-coupling limit. 

\subsection{Magnetic susceptibility}

The dynamical vertex suppresses the unphysical divergence of the RPA and predicts a stable spin symmetric solution for all temperatures and interaction strengths. We now check this conclusion on the magnetic susceptibility in the FLEX and RRPA. Since only the limit to strong coupling matters, it is sufficient to use the polar approximation. We evaluate the magnetic solution only to the linear order in the magnetic field. 

 Notice that the singlet electron-hole bubble is an even function of the magnetic field $h$; thus, its derivative with respect to $h$ does not enter the linear susceptibility. The divergence in the magnetic susceptibility is controlled by the magnetic Kondo scale  $a_{m}= \chi_{0}/\chi$ with the magnetic susceptibility of the non-interacting system $\chi_{0}$. 

Using the symmetry of the magnetic solution, namely 
\begin{subequations}
\begin{align}
\frac{\partial X}{\partial h} &\equiv \frac{\partial X_{+}}{\partial h} = - \frac{\partial X_{-}}{\partial h}\,,
\\
\frac{\partial Y}{\partial h} &\equiv \frac{\partial Y_{+}}{\partial h} = - \frac{\partial Y_{-}}{\partial h}\,,
\end{align}
\end{subequations}
we obtain the necessary derivatives of the real and imaginary parts of the self-energy
\begin{subequations}
\begin{align}
\frac{\partial X}{\partial h} &= \left\{\left[\left(1 + \frac U2\pi\chi\right)  + \frac{\partial X}{\partial h}\right]\frac{\partial}{\partial \omega}  + \frac{\partial Y}{\partial h} \frac{\partial}{\partial Y}\right\}
\nonumber \\
& \times\frac{l_{\Omega}\left(\omega + X\right) - A\left(1 + Y\right)}{\left(\omega + X\right)^{2} + \left(1 + Y\right)^{2}}\,,
\\
\frac{\partial Y}{\partial h} &= -\left\{\left[\left(1 + \frac U2\pi\chi\right)  + \frac{\partial X}{\partial h}\right]\frac{\partial}{\partial \omega}   + \frac{\partial Y}{\partial h} \frac{\partial}{\partial Y}\right\}
\nonumber\\
&\times L\frac{1 + Y}{\left(\omega + X\right)^{2} + \left(1 + Y\right)^{2}} \,.
\end{align}
\end{subequations}

We skipped the frequency variables in functions that are not differentiated with respect to frequency.  

We further introduce the following notation 
\begin{subequations}
\begin{align}
R &=  \frac{S^{2} - T^{2}}{\left(S^{2} + T^{2}\right)^{2}}\,, 
\\
P &= \frac{2 S T}{\left(S^{2} + T^{2}\right)^{2}} \,.
\end{align}
\end{subequations}
%The equations to solve are
%%
%\begin{subequations}
%\begin{align}
%\frac{\partial X}{\partial h} &= \left[\left(1 + \frac U2\pi\chi\right)  + \frac{\partial X}{\partial h}\right] \left(AP - l_{\Omega}R\right) -  \frac{\partial Y}{\partial h}\left(l_{\Omega}P + AR\right) 
%\\
%\frac{\partial Y}{\partial h} &=\left[\left(1 + \frac U2\pi\chi\right)  + \frac{\partial X}{\partial h}\right] LP -  \frac{\partial Y}{\partial h}LR
%\end{align}
%\end{subequations}
%
The solution for the derivative of the dynamical self-energy is
\begin{subequations}
\begin{align}
\frac{\partial X}{\partial h} &= \left(1 + \frac U2\pi\chi\right)
\nonumber \\ &\qquad
\times\left[\frac{\left(1 + LR\right)}{\left(1 + LR\right)^{2} + l_{\Omega}LP^{2}  - AP} - 1\right] \,,
\\
\frac{\partial Y}{\partial h} &= \left(1 + \frac U2\pi\chi\right)\frac{LP}{\left(1 + LR\right)^{2} + lLP^{2}  - AP} \,.
\end{align}
\end{subequations}
Inserting the representations for $R$ and $P$ into this solution, we obtain
\begin{widetext}
\begin{subequations}\label{eq:XY-dh}
\begin{multline}
\frac{\partial X}{\partial h} = \left(1 + \frac U2\pi\chi\right)\left[ - 1 \phantom{\frac12}
\right. \\ \left.
+\ \frac{\left(S^{2} + T^{2}\right)^{2}\left[\left(S^{2} + T^{2}\right)^{2} + L\left(S^{2} - T^{2}\right)\right]}{\left[\left(S^{2} + T^{2}\right)^{2} + L\left(S^{2} - T^{2}\right)\right]\left[\left(S^{2} + T^{2}\right)^{2} + l_{\Omega}\left(S^{2} - T^{2}\right) - 2 AST\right] + 2LST\left[2l_{\Omega}ST + A\left(S^{2} - T^{2}\right)\right]}
\right] \,,
\end{multline}
\begin{multline}
\frac{\partial Y}{\partial h} = \left(1 + \frac U2\pi\chi\right)\
\\ 
\times \frac{2LST\left(S^{2} + T^{2}\right)^{2}}{\left[\left(S^{2} + T^{2}\right)^{2} + L\left(S^{2} - T^{2}\right)\right]\left[\left(S^{2} + T^{2}\right)^{2} + l_{\Omega}\left(S^{2} - T^{2}\right) - 2 AST\right] + 2LST\left[2l_{\Omega}ST + A\left(S^{2} - T^{2}\right)\right]}\,.
\end{multline}
\end{subequations}

The magnetic susceptibility at zero temperature is
\begin{multline}
\pi\chi = 
2\frac{\partial}{\partial h} \int_{-\infty}^{0}d\omega \frac{1 + Y_{+}(\omega)}{\left(\omega + \bar{h} + X_{+}(\omega)\right)^{2} + \left(1 + Y_{+}(\omega)\right)^{2}}  
\\
=   2\left(1 + \frac U2\pi\chi\right)\int_{-\infty}^{0}d\omega\left\{\left(1 + \frac{\partial X}{\partial\bar{h}}\right)\frac{\partial}{\partial \omega} + \frac{\partial Y}{\partial\bar{h}}\frac{\partial}{\partial Y}\right\} \frac{1 + Y}{\left(\omega + X\right)^{2} + \left(1 + Y\right)^{2}} \,.
\\
%= - 2\left(1 + \frac U2\pi\chi\right)\int_{-\infty}^{0}d\omega\left[\left(1 + \frac{\partial X}{\partial\bar{h}}\right)\frac{2ST}{\left(S^{2} + T^{2}\right)^{2}} - \frac{\partial Y}{\partial\bar{h}}\frac{S^{2} - T^{2}}{\left(S^{2} + T^{2}\right)^{2}}\right]\,,
\end{multline}
Inserting Eqs.~\eqref{eq:XY-dh} into the above expression, we obtain an explicit solution
\begin{align}\label{eq:chi-implicit}
\frac{\pi\chi}{2 + U\pi\chi} &=  - 2\int_{-\infty}^{0}d\omega 
\frac{S(\omega)T(\omega)}{\left[\left(S(\omega)^{2} + T(\omega)^{2}\right)^{2} + \left(L(\omega) + l_{\Omega}\right)\left(S(\omega)^{2} - T(\omega)^{2}\right) + l_{\Omega}L(\omega)\right] 
 - 2 A(\omega)S(\omega)T(\omega)}\,. %\nonumber
 \end{align}
 \end{widetext}
%If we denote 
%\begin{equation}\label{eq:I-susceptibility}
%I =  2\int_{0}^{\infty}d\omega 
%%
%\frac{S(\omega)T(\omega)}{\left[\left(S(\omega)^{2} + T(\omega)^{2}\right)^{2} + \left(L(\omega) + l_{\Omega}\right)\left(S(\omega)^{2} - T(\omega)^{2}\right) + l_{\Omega}L(\omega)\right] 
% - 2 A(\omega)S(\omega)T(\omega)} 
% \end{equation}then the susceptibility is 

\begin{figure}
\includegraphics[width=75mm]{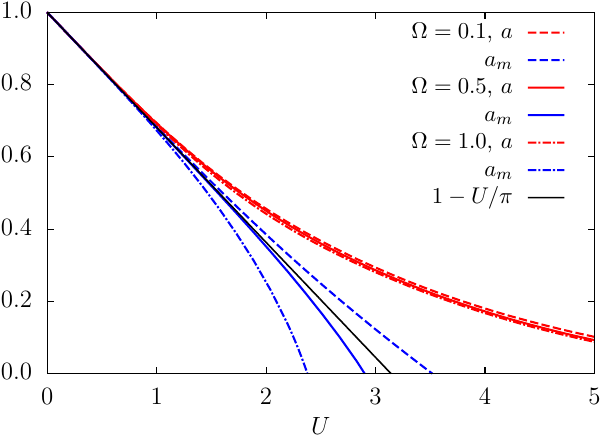}
\caption{The Kondo scale $a$ from the dynamical vertex, red curves, and the magnetic Kondo scale $a_{m}= \chi_{0}/\chi$, blue curves,  plotted for different values of the cutoff $\Omega$; $\chi_{0},\chi$ are zero-temperature bare and full susceptibilities calculated from Eq.~\eqref{eq:chi-implicit}. The black solid line is the Hartree solution. Notice that only the magnetic scale depends on the cutoff.  \label{fig:susc-Omega}}
\end{figure}

We further rewrite the equation for the magnetic susceptibility to 
\be
\pi\chi = \frac{2I}{1 - UI }\,,
\ee
where $I$ is the integral from the right-hand side of Eq.~\eqref{eq:chi-implicit}. 
The Kondo scale from the magnetic susceptibility is 
\be
a_{m} = 1 - UI \,.
\ee
We plotted the two Kondo scales $a$ and $a_{m}$ in Fig.~\ref{fig:susc-Omega} as functions of the interaction strength. We compared their values for a few choices of the cutoff frequency $\Omega$. The smaller the cutoff frequency, the sooner the strong-coupling regime with a critical point is reached. But whatever cutoff frequency is chosen, there is always a finite critical interaction $U_{c}$ at which $a_{m}=0$ and the magnetic susceptibility diverges. The Kondo scale from the dynamical vertex only slightly depends on the cutoff and reaches its critical value $a=0$ at $U_{c}= \infty$.

The susceptibility from the algebraic equation for the self-energy differs from the result derived from the convolutive Schwinger-Dyson equation. It asymptotically approaches, however,  the exact FLEX value in the strong-coupling limit where it has the following explicit representation expressed in terms of the logarithm of the Kondo scale $l = |\ln a|$, see Appendix B2,    
\begin{subequations}\label{eq:chi-SC}
\begin{align}
\frac{\pi\chi}{2 + U\pi\chi} &=- \frac 1{\sqrt{l}} P\int_{0}^{1}dx \frac{\sqrt{1 - x^{2}}}{x - \displaystyle{\frac{\pi}{4l}}\sqrt{1 - x^{2}}}\,.
 \end{align}
\begin{align}
\chi_{SC} &= \frac2\pi \frac{\ln\left(\displaystyle{\frac4\pi}|\ln a|\right)}{1 - \displaystyle{\frac {3\pi}{4} \ln\left(\displaystyle{\frac4\pi}|\ln a|\right)}} \,,
\end{align}
\end{subequations}
It is negative in the limit $a\to0$. Inserting the asymptotic solution for the Kondo scale, Eq.~\eqref{eq:a-asymptotic}, we obtain a critical interaction $U_{c}\approx 3.3\Delta$ for the magnetic instability.  Since the polar approximation becomes asymptotically exact in the strong-coupling limit, the full susceptibility calculated from the conserving vertex of the convolutive solution becomes infinite at a finite value of the interaction strength. Therefore, the spin-symmetric state of the self-consistent approximations becomes unstable at a finite interaction strength $U_{c}<\infty$ when determined from the conserving vertex.  The critical interaction for the magnetic instability obtained from the algebraic approximation sets a lower bound for the actual instability. What holds exactly is that the magnetic susceptibility of the spin-symmetric solution becomes negative in the strong-coupling regime. 

%We, hence, surprisingly found that the magnetic susceptibility diverges and signals magnetic instability of the self-consistent approximations, ladder, RRPA, FLEX, and ald also parquets, of the spin-symmetric state at a finite interaction strength of the SIAM.

\section{Conclusions: Stability of approximate solutions}

The Baym-Kadanoff scheme with the Luttinger-Ward functional is the canonical way to build thermodynamically consistent and conserving approximations of correlated fermions. It is fully self-consistent in the one-electron functions, self-energy and Green function that are uniquely defined. Two-particle and higher-order functions can be derived only indirectly, and there are two ways to connect them with the one-particle self-energy. The straightforward way to introduce the two-particle vertex is the Schwinger-Dyson equation that singles out the two-particle contribution to the self-energy. This equation represents the complete solution of the perturbation theory and is, hence, diagrammatically fully controlled. The conserving character of the solutions is, however, not automatically guaranteed by the Schwinger-Dyson equation. Another equation, Ward identity, matching the two-particle irreducible vertex with the self-energy must be obeyed. It is an integral equation with a functional derivative, Eq.~\eqref{eq:SDE-derivative}, that is equivalent to the full solution of the Schwinger field theory. The dynamical and conserving vertex are not equal in approximate solutions. This difference becomes a severe problem for determining the stability of solutions and identifying the true critical behavior.     

We used the simplest dynamical approximation with critical behavior to demonstrate the generic deficiency of approximate solutions within the Baym-Kadanoff construction: the dynamical vertex from the Schwinger-Dyson equation and the conserving one do not lead to the same critical behavior. The dynamical vertex is diagrammatically derivable and controllable and sets the quality of the approximations. The conclusion about the stability of the spin-symmetric state is unreliable when judged only from the dynamical vertex. The conserving vertex must be used to confirm the conclusion drawn from the dynamical vertex.  The conserving vertex contains, however, additional  Feynman diagrams beyond those from the dynamical one that are difficult to control \cite{Levine:1968aa,Janis:1998aa}.  That is why the conserving vertex is rarely used to determine the stability of high-temperature solutions. 
%The renormalized conserving approximations, even if they include advanced constructions containing parquet equations, dynamical-cluster, or dual fermions, always lead to a mismatch of the critical behavior of the dynamical and conserving vertices.   

The existence of different critical points in the dynamical and conserving vertex is an artifact of approximations. This difference is standardly resolved by picking the divergence in the conserving vertex that comes first and by identifying it as the real one. The conserving vertex does not, however, enter any self-consistent equation, and hence, it is not forced to comply with restrictions such as the Mermin-Wagner theorem.  The critical behavior of the conserving vertex is reliable only at sufficiently high temperatures where quantum fluctuations do not play an important role and a thermodynamic mean-field description is sufficient. Decreasing the temperature increases the impact of quantum fluctuations, and the lower the critical temperature, the less reliable the critical behavior of the conserving vertex is. The low-temperature critical behavior is significantly affected by quantum fluctuations and should be determined from the dynamical vertex. It is the case of all low-dimensional systems, $d<3$. We demonstrated it explicitly in an example of the SIAM where the critical point of the dynamical vertex is in accord with the exact solution at $T=0$ and $U=\infty$ while the conserving vertex leads to a spurious magnetic instability at intermediate coupling in a similar way as the Hartree static solution. We hence showed that the dynamical vertex better reproduces the critical behavior than the one of the conserving vertex in the low-dimensional systems with non-integrable singularity. 
%Even if we successfully suppress spurious transitions in the conserving vertex, we still do not reach the complete thermodynamic consistency unless the conserving vertex shares the critical point with the dynamical one. 

%We recently used approximations with a two-particle self-consistency and parquet equations to make the critical behavior in dynamical and conserving vertices identical, but only with unrenormalized, Hartree one-particle Green functions \cite{Janis:2019aa}. It is not, however, a thermodynamically consistent approximation with a Luttinger-Ward generating functional.

Thermodynamically consistent approximations must lead to unique critical behavior so that scaling arguments and the renormalization group can be applied beyond the mean field in spatial dimensions $d<4$. It means that the characteristic scale vanishing at the critical point must be at least qualitatively the same whether determined from the dynamical vertex (spectral function), conserving vertex (magnetization), or the specific heat. The critical behavior in the specific heat is related to the self-energy and the pole in the dynamical vertex, but only if the full one-particle self-consistency is used. Matching the critical behavior in the specific heat, the dynamical and conserving vertices is possible only with full one- and two-particle self-consistency. It is, however, unreachable with the existing approximate constructions. New ways to match quantum dynamics and thermodynamic criticality in the  Baym-Kadanoff scheme must be devised. Although complete consistency seems out of reach, at least, qualitative consistency between the critical behavior in thermodynamic functions should be aimed at.          

\section*{Acknowledgment}
VP was supported by Grant No. 23-05263K of the Czech Science Foundation. \v{S}K was supported by the project QM4ST No. CZ.02.01.01/00/22-008/0004572, co-funded by the ERDF as part of the MEYS. 

\appendix

\section{Low-temperature integrals in the polar approximation of SIAM}

We replace the full frequency-dependent bubble with its leading low-frequency terms   
\be
\phi(x) \doteq \phi_{0} - i\phi^{\prime}x \,,
\ee
where the real frequency $x\to x + i0^{+}$ is the limit from the upper complex half-plane. Since the SIAM is a local Fermi liquid, $\phi^{\prime} = \pi\rho_{0}^{2} = 1/\pi\Delta^{2}$ independently of the interaction strength. This low-frequency asymptotics gives the exact leading logarithmic divergence. When we use this approximation away from the critical point, we must introduce a frequency cutoff $\Omega$. It will affect only non-universal corrections to the logarithmic divergence in the SDE. The cutoff will be optimized to fit best the Fermi-liquid properties in the weak-coupling limit. 

The two-particle vertex in the SDE in this approximation is
%\begin{subequations}
\begin{align}
\Im \left[\frac{\phi(x) }{1 + U\phi(x) }\right] &= \frac{-\phi^{\prime}x}{a^{2} + U^{2}\phi^{\prime 2}x^{2}}\,,
\\
\Re  \left[\frac{\phi(x) }{1 + U\phi(x) }\right] &= \frac{a\phi_{0}}{a^{2} + U^{2}\phi^{\prime 2}x^{2}}\,.
\end{align}
%\end{subequations}

The Kondo limit with vanishing of the Kondo scale $a \to 0$ becomes critical at charge and spin symmetric case at zero temperature.  We can replace the integrals over the bosonic and fermionic variables in the two-particle vertex in the low-temperature critical region with 
%\begin{subequations}
\begin{multline}
\int_{-\Omega}^{\Omega}\frac{dx}{\pi} b(x)\Im \left[\frac{\phi(x) }{1 + U\phi(x) }\right] 
\\
\doteq  - \frac{1}{U^{2}\pi \phi^{\prime}}\ln\sqrt{1 + \frac{U^{2}\phi^{\prime 2}\Omega^{2}}{a^{2}}}\,, 
\end{multline}
\begin{multline}
\int_{-\Omega}^{\Omega}\frac{dx}{\pi} f(\omega + x) \Re \left[\frac{\phi(x) }{1 + U\phi(x) }\right] 
\\
\doteq  - \frac{\phi_{0}}{U\pi \phi^{\prime}} \arctan\left(\frac{U\phi^{\prime}\omega}{a}\right)\,,
\end{multline}
\begin{multline}
\int_{-\Omega}^{\Omega}\frac{dx}{\pi} \left(b(x) + f(\omega + x)\right)\Im \left[\frac{\phi(x) }{1 + U\phi(x) }\right] 
\\
\doteq  - \frac{1}{U^{2}\pi \phi^{\prime}}\ln\sqrt{1 + \frac{U^{2}\phi^{\prime 2}\omega^{2}}{a^{2}}}\,.
\end{multline}
%\end{subequations}  
We assume $|\omega|<\Omega$ when the approximation is applied outside the critical region. We utilized the symmetry of the dynamical self-energy at half filling with $\Re\Sigma(\omega) = - \Re\Sigma(-\omega)$ and $\Im\Sigma(\omega) =  \Im\Sigma(-\omega)$ for real frequencies. We next use  
$U\phi_{0} = \left(a - 1\right)$.  

\section{Asymptotic limits of the polar approximation}

\subsection{Weak-coupling - integration cutoff}
\label{sec:Weak}

It is necessary to choose the frequency cutoff since it is a free parameter of the approximation outside the critical region where the Kondo scale in weak coupling approaches one. We choose the cutoff to fit the derivative $dX(\omega)/d\omega$  at $\omega=0$ of the full solution to the second order of the interaction strength $U$.  In this way, we reproduce the Fermi-liquid properties best. We use the bare propagator and the bare electron-hole bubble in the weak-coupling limit
\be
\Re G(\omega) + i\Im G(\omega) = \frac{\omega - i}{\omega^{2} + 1}
\ee
and
%\begin{subequations}
\begin{multline}
\Re \phi(x) =  \frac 1\pi\int_{-\infty}^{0}\frac{d\omega}{\omega^{2} + 1}
\\
\times\left[\frac{\omega + x}{(\omega + x)^{2} + 1} + \frac{\omega - x}{(\omega - x)^{2} + 1}\right] 
\\
= -\frac{4\arctan(x) - x \ln\left(1 + x^{2}\right)}{x(4 + x^{2})\pi}\,,
\end{multline}
\begin{multline}
\Im \phi(x) =  - \frac 1\pi\int_{-\infty}^{0}\frac{d\omega}{\omega^{2} + 1}
\\
\times\left[\frac{1}{(\omega + x)^{2} + 1} - \frac{1}{(\omega - x)^{2} + 1}\right] 
\\
= - 2 \frac{x\arctan(x) + \ln\left(1 + x^{2}\right)}{x(4 + x^{2})\pi}\,.
\end{multline}
%\end{subequations}

The leading contribution to the derivative of the dynamical self-energy at the Fermi level at zero temperature is
\begin{multline}
\left.\frac{d X(\omega)}{d\omega}\right|_{\omega=0} =   \frac{U^{2}\Im G(0)\phi(0)}{\pi}
\\
- U^{2} \int_{-\infty}^{0}\frac{dx}{\pi}\left[\frac{d \Re G(x)}{dx} \Im\phi(x) + \frac{d \Im G(x)}{dx} \Re\phi(x)\right]  
\\
= \frac{U^{2}}{\pi^{2}} +  \frac{2U^{2}}{\pi^{2}}\int_{-\infty}^{0}\frac{dx}{\left(1 + x^{2}\right)^{2}}
\\
\times\left[\left(1 - x^{2}\right)\frac{x\arctan(x) + \ln\left(1 + x^{2}\right)}{x(4 + x^{2})}
\right. \\ \left.  +\  \frac{4\arctan(x) - x \ln\left(1 + x^{2}\right)}{(4 + x^{2})}\right] 
\approx 0.533  \frac{U^{2}}{\pi^{2}}
\end{multline}
%The same derivative from the polar approximation is

%This value leads to $\alpha\approx 0.15$.
%The result is
%\be
%\left.\frac{d\Re\Sigma(\omega)}{d\omega}\right|_{\omega=0} \approx - 1.45\frac{U^{2}}{\pi^{2}}
%\ee
%The same derivative from the polar approximation is
%\be
%\left.\frac{d\Re\Sigma(\omega)}{d\omega}\right|_{\omega=0} = - \frac{U^{2}}{2\pi^{2}}\left(2 - \Omega^{2}\right)
%\ee
%Equaling the two results leads to the frequency cutoff

%The susceptibility in the physical units is
%\be
%\chi \doteq \frac{2\rho_{0}}{1 - U\rho_{0}\left(1 - \displaystyle{\frac{5U^{2}\rho_{0}^{2}}3}\right)} \left(1 - \displaystyle{\frac{5U^{2}\rho_{0}^{2}}3}\right)
%\ee

%\begin{multline}
%\frac{\pi\chi}{2 +  U\pi\chi} = 1 - U(1 - a)\int_{-\infty}^{0}\frac{d\omega}{1 + U^{2}\omega^{2}}  \frac{1 + Y(\omega)}{\left[\left(\omega + X(\omega)\right)^{2} + \left(1 + Y(\omega)\right)^{2}\right]^{3}}\left\{
%2\left(1 + Y(\omega)\right)\left(\omega + X(\omega)\right) 
%\right. \\ \left.
%+\ U\omega\left[\left(1 + Y(\omega)\right)^{2} - \left(\omega + X(\omega)\right)^{2}\right] \
%\right\}
%\end{multline}
% We recall that the interaction strength was rescaled to $U/\pi$ in the above formulas and $1 - a = U$ in the weak-coupling limit. The leading-order correction to the Hartree susceptibility in the weak-coupling limit is
%\be
%\frac{\pi\chi}{2 +  U\pi\chi} = 1 - U^{2}\int_{-\infty}^{\infty}d\omega\frac{2\omega}{\left(\omega^{2} + 1\right)^{3}} = 1 + \frac{U^{2}}2
%\ee
%It gives 
%\be
%\pi\chi = \frac{2 + U^{2}}{2 - U\left(2 + U^{2}\right)}
%\ee

\subsection{Strong-coupling - magnetic susceptibility}
\label{sec:Strong}

Using functions $S(\omega)$, $T(\omega)$, and $Z(\omega)$ in the Green function, defined in Eqs.~\eqref{eq:STZ-Def}  we obtain the following asymptotic equations
%\begin{subequations}
\begin{align}
S(\omega) &=  \omega -  \frac{lS(\omega) - \frac\pi2\mathrm{sign}(\omega) T(\omega)}{Z(\omega)^{2}}\,, 
\\
T(\omega) &=  1 + \frac{l T(\omega)}{Z(\omega)^{2}}\,.
\end{align}
%\end{subequations}
where $l = |\ln a|$. Their solution is
%\begin{subequations}
\begin{align}
T(\omega) &=  \frac{Z(\omega)^{2}}{Z(\omega)^{2} - l} \,,
\\
S(\omega) &= \frac{\omega Z(\omega)^{2}}{Z(\omega)^{2} + l}\,.
\end{align}
%\end{subequations}

The equation for $Z(\omega)^{2}$ in the strong-coupling reduces to
\begin{multline}\label{eq:Z2-strong}
\left[Z(\omega)^{4} - l^{2}\right]^{2} = Z(\omega)^{2}
\\
\times\left[\left(Z(\omega)^{2} + l\right)^{2} + \omega^{2}\left(Z(\omega)^{2} - l\right)^{2}\right]\,.
\end{multline}
If we substitute $Z(\omega) = \zeta(\omega)\sqrt{l}$, then
\be
 l = \zeta(\omega)^{2}\left[\frac 1{\left(\zeta(\omega)^{2} - 1\right)^{2}} + \frac {\omega^{2}}{\left(\zeta(\omega)^{2} + 1\right)^{2}}\right]\,.
\ee  

In the next step, we represent the strong-coupling solution in the leading order of small parameter $1/\sqrt{l}$
\be
\zeta(\omega)^{2} = 1 + \frac{\alpha(\omega)}{\sqrt{l}} \,,
\ee
where $\alpha(\omega) \ll \sqrt{l}$.

The equation for function $\alpha(\omega)$ from Eq.~\eqref{eq:Z2-strong} is
\be
 l = \frac {l}{\alpha(\omega)^{2}} + \frac {\omega^{2}}{4}\,.
\ee  
Introducing a new variable $x= \omega/2\sqrt{l}$ we obtain 
\be
\alpha(\omega)^{2} = \frac 1{1 - x^{2}}\,.
\ee

The solution in the leading order of $1/\sqrt{l}$ is
%\begin{subequations}
\begin{align}
Z(x) &= \sqrt{l}\,,
\\
T(x) &= \sqrt{l}\sqrt{1 - x^{2}}\,, 
\\
S(x) &=  x \sqrt{l}\,,
\end{align}
%\end{subequations}
from which we obtain the equation for the Kondo scale in the strong-coupling limit, Eq.~\eqref{eq:a-strong-equation}.

%The Kondo scale in the strong-coupling limit is determined from the following equation
%\be
%1 = \frac{4U}{\pi\sqrt{|\ln a|}}\int_{0}^{1}dx x\sqrt{1 - x^{2}} =  \frac{4U}{3\pi\sqrt{|\ln a|}} \,.
%\ee
%It means 
%\be
%a = \exp\left\{- \frac{16U^{2}}{9\pi^{2}}\right\}\,,
%\ee
%the result reported in the main text in Eqs.~(4),~(5).

We obtain the magnetic susceptibility in the strong coupling Kondo limit by inserting the above solution into Eq.~\eqref{eq:chi-implicit} 
\begin{widetext}
\begin{align}\label{eq:chi-Appendix}
\frac{\pi\chi}{2 + U\pi\chi} &=- 2\int_{-\infty}^{0}d\omega 
\frac{S(\omega)T(\omega)}{\left[\left(S(\omega)^{2} + T(\omega)^{2}\right)^{2} +  2l\left(S(\omega)^{2} - T(\omega)^{2}\right) + l^{2}\right] 
 + \pi S(\omega)T(\omega)} 
\nonumber \\
&= \frac 1{\sqrt{l}} P\int_{0}^{1}dx \frac{\sqrt{1 - x^{2}}}{x - \displaystyle{\frac{\pi}{4l}}\sqrt{1 - x^{2}}}\,,
 \end{align}
being Eq.~\eqref{eq:chi-SC}.
 \end{widetext}

%\bibliographystyle{apsrev}
%\bibliography{../../../../BibTeX/parquets_MB,../../../../BibTeX/Impurity_solver-1,../../../../BibTeX/FLEX-Kondo-Supplement}

\begin{thebibliography}{42}
\expandafter\ifx\csname natexlab\endcsname\relax\def\natexlab#1{#1}\fi
\expandafter\ifx\csname bibnamefont\endcsname\relax
  \def\bibnamefont#1{#1}\fi
\expandafter\ifx\csname bibfnamefont\endcsname\relax
  \def\bibfnamefont#1{#1}\fi
\expandafter\ifx\csname citenamefont\endcsname\relax
  \def\citenamefont#1{#1}\fi
\expandafter\ifx\csname url\endcsname\relax
  \def\url#1{\texttt{#1}}\fi
\expandafter\ifx\csname urlprefix\endcsname\relax\def\urlprefix{URL }\fi
\providecommand{\bibinfo}[2]{#2}
\providecommand{\eprint}[2][]{\url{#2}}

\bibitem[{\citenamefont{Wiegmann and Tsvelick}(1983)}]{Wiegmann:1983aa}
\bibinfo{author}{\bibfnamefont{P.~B.} \bibnamefont{Wiegmann}} \bibnamefont{and}
  \bibinfo{author}{\bibfnamefont{A.~M.} \bibnamefont{Tsvelick}},
  \bibinfo{journal}{Journal of Physics C: Solid State Physics}
  \textbf{\bibinfo{volume}{16}}, \bibinfo{pages}{2281} (\bibinfo{year}{1983}).

\bibitem[{\citenamefont{Tsvelick and Wiegmann}(1983)}]{Tsvelick:1983aa}
\bibinfo{author}{\bibfnamefont{A.}~\bibnamefont{Tsvelick}} \bibnamefont{and}
  \bibinfo{author}{\bibfnamefont{P.}~\bibnamefont{Wiegmann}},
  \bibinfo{journal}{Advances in Physics} \textbf{\bibinfo{volume}{32}},
  \bibinfo{pages}{453 } (\bibinfo{year}{1983}).

\bibitem[{\citenamefont{Wilson}(1975)}]{Wilson:1975aa}
\bibinfo{author}{\bibfnamefont{K.~G.} \bibnamefont{Wilson}},
  \bibinfo{journal}{Reviews of Modern Physics} \textbf{\bibinfo{volume}{47}},
  \bibinfo{pages}{773} (\bibinfo{year}{1975}).

\bibitem[{\citenamefont{Andrei et~al.}(1983)\citenamefont{Andrei, Furuya, and
  Lowenstein}}]{Andrei:1983aa}
\bibinfo{author}{\bibfnamefont{N.}~\bibnamefont{Andrei}},
  \bibinfo{author}{\bibfnamefont{K.}~\bibnamefont{Furuya}}, \bibnamefont{and}
  \bibinfo{author}{\bibfnamefont{J.~H.} \bibnamefont{Lowenstein}},
  \bibinfo{journal}{Reviews of Modern Physics} \textbf{\bibinfo{volume}{55}},
  \bibinfo{pages}{331} (\bibinfo{year}{1983}).

\bibitem[{\citenamefont{Lieb and Wu}(1968)}]{Lieb:1968aa}
\bibinfo{author}{\bibfnamefont{E.~H.} \bibnamefont{Lieb}} \bibnamefont{and}
  \bibinfo{author}{\bibfnamefont{F.~Y.} \bibnamefont{Wu}},
  \bibinfo{journal}{Physical Review Letters} \textbf{\bibinfo{volume}{20}},
  \bibinfo{pages}{1445} (\bibinfo{year}{1968}).

\bibitem[{\citenamefont{Essler et~al.}(2005)\citenamefont{Essler, Frahm,
  G{\"o}hmann, Kl{\"u}mper, and Korepin}}]{Essler:2005aa}
\bibinfo{author}{\bibfnamefont{F.~H.~L.} \bibnamefont{Essler}},
  \bibinfo{author}{\bibfnamefont{H.}~\bibnamefont{Frahm}},
  \bibinfo{author}{\bibfnamefont{F.}~\bibnamefont{G{\"o}hmann}},
  \bibinfo{author}{\bibfnamefont{A.}~\bibnamefont{Kl{\"u}mper}},
  \bibnamefont{and} \bibinfo{author}{\bibfnamefont{V.~E.}
  \bibnamefont{Korepin}}, \emph{\bibinfo{title}{The One-Dimensional Hubbard
  Model}} (\bibinfo{publisher}{Cambridge University Press},
  \bibinfo{address}{Cambridge}, \bibinfo{year}{2005}).

\bibitem[{\citenamefont{Hewson}(1993)}]{Hewson:1993aa}
\bibinfo{author}{\bibfnamefont{A.~C.} \bibnamefont{Hewson}},
  \emph{\bibinfo{title}{The Kondo Problem to Heavy Fermions}},
  vol.~\bibinfo{volume}{2} of \emph{\bibinfo{series}{Cambridge Studies in
  Magnetism}} (\bibinfo{publisher}{Cambridge University Press},
  \bibinfo{address}{Cambridge}, \bibinfo{year}{1993}).

\bibitem[{\citenamefont{Baym and Kadanoff}(1961)}]{Baym:1961aa}
\bibinfo{author}{\bibfnamefont{G.}~\bibnamefont{Baym}} \bibnamefont{and}
  \bibinfo{author}{\bibfnamefont{L.~P.} \bibnamefont{Kadanoff}},
  \bibinfo{journal}{Physical Review} \textbf{\bibinfo{volume}{124}},
  \bibinfo{pages}{287} (\bibinfo{year}{1961}).

\bibitem[{\citenamefont{Baym}(1962)}]{Baym:1962aa}
\bibinfo{author}{\bibfnamefont{G.}~\bibnamefont{Baym}},
  \bibinfo{journal}{Physical Review} \textbf{\bibinfo{volume}{127}},
  \bibinfo{pages}{1391} (\bibinfo{year}{1962}).

\bibitem[{\citenamefont{Dominicis}(1963)}]{DeDominicis:1963aa}
\bibinfo{author}{\bibfnamefont{C.~D.} \bibnamefont{Dominicis}},
  \bibinfo{journal}{Journal of Mathematical Physics}
  \textbf{\bibinfo{volume}{4}}, \bibinfo{pages}{255} (\bibinfo{year}{1963}).

\bibitem[{\citenamefont{Dominicis and
  Martin}(1964{\natexlab{a}})}]{DeDominicis:1964aa}
\bibinfo{author}{\bibfnamefont{C.~D.} \bibnamefont{Dominicis}}
  \bibnamefont{and} \bibinfo{author}{\bibfnamefont{P.~C.}
  \bibnamefont{Martin}}, \bibinfo{journal}{Journal of Mathematical Physics}
  \textbf{\bibinfo{volume}{5}}, \bibinfo{pages}{14}
  (\bibinfo{year}{1964}{\natexlab{a}}).

\bibitem[{\citenamefont{Dominicis and
  Martin}(1964{\natexlab{b}})}]{DeDominicis:1964ab}
\bibinfo{author}{\bibfnamefont{C.~D.} \bibnamefont{Dominicis}}
  \bibnamefont{and} \bibinfo{author}{\bibfnamefont{P.~C.}
  \bibnamefont{Martin}}, \bibinfo{journal}{Journal of Mathematical Physics}
  \textbf{\bibinfo{volume}{5}}, \bibinfo{pages}{31}
  (\bibinfo{year}{1964}{\natexlab{b}}).

\bibitem[{\citenamefont{Jani{\v s}}(1998)}]{Janis:1998aa}
\bibinfo{author}{\bibfnamefont{V.}~\bibnamefont{Jani{\v s}}},
  \bibinfo{journal}{Journal of Physics: Condensed Matter}
  \textbf{\bibinfo{volume}{10}}, \bibinfo{pages}{2915} (\bibinfo{year}{1998}).

\bibitem[{\citenamefont{Jani{\v s}
  et~al.}(2017{\natexlab{a}})\citenamefont{Jani{\v s}, Kauch, and
  Pokorn{\'y}}}]{Janis:2017aa}
\bibinfo{author}{\bibfnamefont{V.}~\bibnamefont{Jani{\v s}}},
  \bibinfo{author}{\bibfnamefont{A.}~\bibnamefont{Kauch}}, \bibnamefont{and}
  \bibinfo{author}{\bibfnamefont{V.}~\bibnamefont{Pokorn{\'y}}},
  \bibinfo{journal}{Physical Review B} \textbf{\bibinfo{volume}{95}},
  \bibinfo{pages}{045108} (\bibinfo{year}{2017}{\natexlab{a}}).

\bibitem[{\citenamefont{Schwinger}(1949)}]{Schwinger:1949aa}
\bibinfo{author}{\bibfnamefont{J.}~\bibnamefont{Schwinger}},
  \bibinfo{journal}{Physical Review} \textbf{\bibinfo{volume}{75}},
  \bibinfo{pages}{651} (\bibinfo{year}{1949}).

\bibitem[{\citenamefont{Martin}(1959)}]{Martin:1959aa}
\bibinfo{author}{\bibfnamefont{P.~C.} \bibnamefont{Martin}},
  \bibinfo{journal}{Physical Review} \textbf{\bibinfo{volume}{115}},
  \bibinfo{pages}{1342} (\bibinfo{year}{1959}).

\bibitem[{\citenamefont{Jani{\v s} et~al.}(2019)\citenamefont{Jani{\v s},
  Zalom, Pokorn{\'y}, and Kl{\'\i}{\v c}}}]{Janis:2019aa}
\bibinfo{author}{\bibfnamefont{V.}~\bibnamefont{Jani{\v s}}},
  \bibinfo{author}{\bibfnamefont{P.}~\bibnamefont{Zalom}},
  \bibinfo{author}{\bibfnamefont{V.}~\bibnamefont{Pokorn{\'y}}},
  \bibnamefont{and} \bibinfo{author}{\bibfnamefont{A.}~\bibnamefont{Kl{\'\i}{\v
  c}}}, \bibinfo{journal}{Physical Review B} \textbf{\bibinfo{volume}{100}},
  \bibinfo{pages}{195114} (\bibinfo{year}{2019}).

\bibitem[{\citenamefont{Georges et~al.}(1996)\citenamefont{Georges, Kotliar,
  Krauth, and Rozenberg}}]{Georges:1996aa}
\bibinfo{author}{\bibfnamefont{A.}~\bibnamefont{Georges}},
  \bibinfo{author}{\bibfnamefont{G.}~\bibnamefont{Kotliar}},
  \bibinfo{author}{\bibfnamefont{W.}~\bibnamefont{Krauth}}, \bibnamefont{and}
  \bibinfo{author}{\bibfnamefont{M.~J.} \bibnamefont{Rozenberg}},
  \bibinfo{journal}{Reviews of Modern Physics} \textbf{\bibinfo{volume}{68}},
  \bibinfo{pages}{13} (\bibinfo{year}{1996}).

\bibitem[{\citenamefont{Lichtenstein and
  Katsnelson}(1998)}]{Lichtenstein:1998aa}
\bibinfo{author}{\bibfnamefont{A.~I.} \bibnamefont{Lichtenstein}}
  \bibnamefont{and} \bibinfo{author}{\bibfnamefont{M.~I.}
  \bibnamefont{Katsnelson}}, \bibinfo{journal}{Physical Review B}
  \textbf{\bibinfo{volume}{57}}, \bibinfo{pages}{6884} (\bibinfo{year}{1998}).

\bibitem[{\citenamefont{Jani{\v s}}(1999)}]{Janis:1999aa}
\bibinfo{author}{\bibfnamefont{V.}~\bibnamefont{Jani{\v s}}},
  \bibinfo{journal}{Physical Review B} \textbf{\bibinfo{volume}{60}},
  \bibinfo{pages}{11345} (\bibinfo{year}{1999}).

\bibitem[{\citenamefont{Rohringer et~al.}(2018)\citenamefont{Rohringer,
  Hafermann, Toschi, Katanin, Antipov, Katsnelson, Lichtenstein, Rubtsov, and
  Held}}]{Rohringer:2018aa}
\bibinfo{author}{\bibfnamefont{G.}~\bibnamefont{Rohringer}},
  \bibinfo{author}{\bibfnamefont{H.}~\bibnamefont{Hafermann}},
  \bibinfo{author}{\bibfnamefont{A.}~\bibnamefont{Toschi}},
  \bibinfo{author}{\bibfnamefont{A.}~\bibnamefont{Katanin}},
  \bibinfo{author}{\bibfnamefont{A.}~\bibnamefont{Antipov}},
  \bibinfo{author}{\bibfnamefont{M.}~\bibnamefont{Katsnelson}},
  \bibinfo{author}{\bibfnamefont{A.}~\bibnamefont{Lichtenstein}},
  \bibinfo{author}{\bibfnamefont{A.}~\bibnamefont{Rubtsov}}, \bibnamefont{and}
  \bibinfo{author}{\bibfnamefont{K.}~\bibnamefont{Held}},
  \bibinfo{journal}{Reviews of Modern Physics} \textbf{\bibinfo{volume}{90}},
  \bibinfo{pages}{025003} (\bibinfo{year}{2018}).

\bibitem[{\citenamefont{Bickers et~al.}(1989)\citenamefont{Bickers, Scalapino,
  and White}}]{Bickers:1989aa}
\bibinfo{author}{\bibfnamefont{N.~E.} \bibnamefont{Bickers}},
  \bibinfo{author}{\bibfnamefont{D.~J.} \bibnamefont{Scalapino}},
  \bibnamefont{and} \bibinfo{author}{\bibfnamefont{S.~R.} \bibnamefont{White}},
  \bibinfo{journal}{Physical Review Letters} \textbf{\bibinfo{volume}{62}},
  \bibinfo{pages}{961} (\bibinfo{year}{1989}).

\bibitem[{\citenamefont{Bickers and Scalapino}(1989)}]{Bickers:1989ab}
\bibinfo{author}{\bibfnamefont{N.~E.} \bibnamefont{Bickers}} \bibnamefont{and}
  \bibinfo{author}{\bibfnamefont{D.~J.} \bibnamefont{Scalapino}},
  \bibinfo{journal}{Annals of Physics} \textbf{\bibinfo{volume}{193}},
  \bibinfo{pages}{206} (\bibinfo{year}{1989}).

\bibitem[{\citenamefont{Bickers and White}(1991)}]{Bickers:1991aa}
\bibinfo{author}{\bibfnamefont{N.~E.} \bibnamefont{Bickers}} \bibnamefont{and}
  \bibinfo{author}{\bibfnamefont{S.~R.} \bibnamefont{White}},
  \bibinfo{journal}{Physical Review B} \textbf{\bibinfo{volume}{43}},
  \bibinfo{pages}{8044} (\bibinfo{year}{1991}).

\bibitem[{\citenamefont{Suhl}(1967)}]{Suhl:1967aa}
\bibinfo{author}{\bibfnamefont{H.}~\bibnamefont{Suhl}},
  \bibinfo{journal}{Physical Review Letters} \textbf{\bibinfo{volume}{19}},
  \bibinfo{pages}{442} (\bibinfo{year}{1967}).

\bibitem[{\citenamefont{Levine and Suhl}(1968)}]{Levine:1968aa}
\bibinfo{author}{\bibfnamefont{M.}~\bibnamefont{Levine}} \bibnamefont{and}
  \bibinfo{author}{\bibfnamefont{H.}~\bibnamefont{Suhl}},
  \bibinfo{journal}{Physical Review} \textbf{\bibinfo{volume}{171}},
  \bibinfo{pages}{567} (\bibinfo{year}{1968}).

\bibitem[{\citenamefont{Hamann}(1969)}]{Hamann:1969aa}
\bibinfo{author}{\bibfnamefont{D.~R.} \bibnamefont{Hamann}},
  \bibinfo{journal}{Physical Review} \textbf{\bibinfo{volume}{186}},
  \bibinfo{pages}{549} (\bibinfo{year}{1969}).

\bibitem[{\citenamefont{Jani{\v s} and Augustinsk{\'y}}(2007)}]{Janis:2007aa}
\bibinfo{author}{\bibfnamefont{V.}~\bibnamefont{Jani{\v s}}} \bibnamefont{and}
  \bibinfo{author}{\bibfnamefont{P.}~\bibnamefont{Augustinsk{\'y}}},
  \bibinfo{journal}{Physical Review B} \textbf{\bibinfo{volume}{75}},
  \bibinfo{pages}{165108} (\bibinfo{year}{2007}).

\bibitem[{\citenamefont{Jani{\v s} and Augustinsk{\'y}}(2008)}]{Janis:2008ab}
\bibinfo{author}{\bibfnamefont{V.}~\bibnamefont{Jani{\v s}}} \bibnamefont{and}
  \bibinfo{author}{\bibfnamefont{P.}~\bibnamefont{Augustinsk{\'y}}},
  \bibinfo{journal}{Physical Review B} \textbf{\bibinfo{volume}{77}},
  \bibinfo{pages}{085106} (\bibinfo{year}{2008}).

\bibitem[{\citenamefont{Jani{\v s}
  et~al.}(2017{\natexlab{b}})\citenamefont{Jani{\v s}, Pokorn{\'y}, and
  Kauch}}]{Janis:2017ab}
\bibinfo{author}{\bibfnamefont{V.}~\bibnamefont{Jani{\v s}}},
  \bibinfo{author}{\bibfnamefont{V.}~\bibnamefont{Pokorn{\'y}}},
  \bibnamefont{and} \bibinfo{author}{\bibfnamefont{A.}~\bibnamefont{Kauch}},
  \bibinfo{journal}{Physical Review B} \textbf{\bibinfo{volume}{95}},
  \bibinfo{pages}{165113} (\bibinfo{year}{2017}{\natexlab{b}}).

\bibitem[{\citenamefont{Haussmann et~al.}(2007)\citenamefont{Haussmann,
  Rantner, Cerrito, and Zwerger}}]{Haussmann:2007aa}
\bibinfo{author}{\bibfnamefont{R.}~\bibnamefont{Haussmann}},
  \bibinfo{author}{\bibfnamefont{W.}~\bibnamefont{Rantner}},
  \bibinfo{author}{\bibfnamefont{S.}~\bibnamefont{Cerrito}}, \bibnamefont{and}
  \bibinfo{author}{\bibfnamefont{W.}~\bibnamefont{Zwerger}},
  \bibinfo{journal}{Physical Review A} \textbf{\bibinfo{volume}{75}},
  \bibinfo{pages}{023610} (\bibinfo{year}{2007}).

\bibitem[{\citenamefont{He and Guo}(2016)}]{He:2016aa}
\bibinfo{author}{\bibfnamefont{Y.}~\bibnamefont{He}} \bibnamefont{and}
  \bibinfo{author}{\bibfnamefont{H.}~\bibnamefont{Guo}},
  \bibinfo{journal}{Physics Letters A} \textbf{\bibinfo{volume}{380}},
  \bibinfo{pages}{2430} (\bibinfo{year}{2016}).

\bibitem[{\citenamefont{Moriya and Ueda}(2000)}]{Moriya:2000aa}
\bibinfo{author}{\bibfnamefont{T.}~\bibnamefont{Moriya}} \bibnamefont{and}
  \bibinfo{author}{\bibfnamefont{K.}~\bibnamefont{Ueda}},
  \bibinfo{journal}{Advances in Physics} \textbf{\bibinfo{volume}{49}},
  \bibinfo{pages}{555} (\bibinfo{year}{2000}).

\bibitem[{\citenamefont{Kemper et~al.}(2010)\citenamefont{Kemper, Maier,
  Graser, Cheng, Hirschfeld, and Scalapino}}]{Kemper:2010aa}
\bibinfo{author}{\bibfnamefont{A.~F.} \bibnamefont{Kemper}},
  \bibinfo{author}{\bibfnamefont{T.~A.} \bibnamefont{Maier}},
  \bibinfo{author}{\bibfnamefont{S.}~\bibnamefont{Graser}},
  \bibinfo{author}{\bibfnamefont{H.-P.} \bibnamefont{Cheng}},
  \bibinfo{author}{\bibfnamefont{P.~J.} \bibnamefont{Hirschfeld}},
  \bibnamefont{and} \bibinfo{author}{\bibfnamefont{D.~J.}
  \bibnamefont{Scalapino}}, \bibinfo{journal}{New Journal of Physics}
  \textbf{\bibinfo{volume}{12}}, \bibinfo{pages}{073030}
  (\bibinfo{year}{2010}).

\bibitem[{\citenamefont{Onari and Kontani}(2012)}]{Onari:2012aa}
\bibinfo{author}{\bibfnamefont{S.}~\bibnamefont{Onari}} \bibnamefont{and}
  \bibinfo{author}{\bibfnamefont{H.}~\bibnamefont{Kontani}},
  \bibinfo{journal}{Physical Review Letters} \textbf{\bibinfo{volume}{109}},
  \bibinfo{pages}{137001} (\bibinfo{year}{2012}).

\bibitem[{\citenamefont{Hirschmeier et~al.}(2015)\citenamefont{Hirschmeier,
  Hafermann, Gull, Lichtenstein, and Antipov}}]{Hirschmeier:2015aa}
\bibinfo{author}{\bibfnamefont{D.}~\bibnamefont{Hirschmeier}},
  \bibinfo{author}{\bibfnamefont{H.}~\bibnamefont{Hafermann}},
  \bibinfo{author}{\bibfnamefont{E.}~\bibnamefont{Gull}},
  \bibinfo{author}{\bibfnamefont{A.~I.} \bibnamefont{Lichtenstein}},
  \bibnamefont{and} \bibinfo{author}{\bibfnamefont{A.~E.}
  \bibnamefont{Antipov}}, \bibinfo{journal}{Physical Review B}
  \textbf{\bibinfo{volume}{92}}, \bibinfo{pages}{144409}
  (\bibinfo{year}{2015}).

\bibitem[{\citenamefont{Re and Rohringer}(2021)}]{Re:2021aa}
\bibinfo{author}{\bibfnamefont{L.~D.} \bibnamefont{Re}} \bibnamefont{and}
  \bibinfo{author}{\bibfnamefont{G.}~\bibnamefont{Rohringer}},
  \bibinfo{journal}{Physical Review B} \textbf{\bibinfo{volume}{104}},
  \bibinfo{pages}{235128} (\bibinfo{year}{2021}).

\bibitem[{\citenamefont{Witt et~al.}(2021)\citenamefont{Witt, van Loon, Nomoto,
  Arita, and Wehling}}]{Witt:2021aa}
\bibinfo{author}{\bibfnamefont{N.}~\bibnamefont{Witt}},
  \bibinfo{author}{\bibfnamefont{E.~G. C.~P.} \bibnamefont{van Loon}},
  \bibinfo{author}{\bibfnamefont{T.}~\bibnamefont{Nomoto}},
  \bibinfo{author}{\bibfnamefont{R.}~\bibnamefont{Arita}}, \bibnamefont{and}
  \bibinfo{author}{\bibfnamefont{T.~O.} \bibnamefont{Wehling}},
  \bibinfo{journal}{Physical Review B} \textbf{\bibinfo{volume}{103}}
  (\bibinfo{year}{2021}).

\bibitem[{\citenamefont{Klett et~al.}(2022)\citenamefont{Klett, Hansmann, and
  Sch{\"a}fer}}]{Klett:2022aa}
\bibinfo{author}{\bibfnamefont{M.}~\bibnamefont{Klett}},
  \bibinfo{author}{\bibfnamefont{P.}~\bibnamefont{Hansmann}}, \bibnamefont{and}
  \bibinfo{author}{\bibfnamefont{T.}~\bibnamefont{Sch{\"a}fer}},
  \bibinfo{journal}{Frontiers in Physics} \textbf{\bibinfo{volume}{10}},
  \bibinfo{pages}{834682} (\bibinfo{year}{2022}).

\bibitem[{\citenamefont{Witt et~al.}(2023)\citenamefont{Witt, Si, Tomczak,
  Held, and Wehling}}]{Witt:2023aa}
\bibinfo{author}{\bibfnamefont{N.}~\bibnamefont{Witt}},
  \bibinfo{author}{\bibfnamefont{L.}~\bibnamefont{Si}},
  \bibinfo{author}{\bibfnamefont{J.~M.} \bibnamefont{Tomczak}},
  \bibinfo{author}{\bibfnamefont{K.}~\bibnamefont{Held}}, \bibnamefont{and}
  \bibinfo{author}{\bibfnamefont{T.}~\bibnamefont{Wehling}},
  \bibinfo{journal}{SciPost Physics} \textbf{\bibinfo{volume}{15}},
  \bibinfo{pages}{197} (\bibinfo{year}{2023}).

\bibitem[{\citenamefont{Li et~al.}(2016)\citenamefont{Li, Wentzell, Pudleiner,
  Thunstr{\"o}m, and Held}}]{Li:2016aa}
\bibinfo{author}{\bibfnamefont{G.}~\bibnamefont{Li}},
  \bibinfo{author}{\bibfnamefont{N.}~\bibnamefont{Wentzell}},
  \bibinfo{author}{\bibfnamefont{P.}~\bibnamefont{Pudleiner}},
  \bibinfo{author}{\bibfnamefont{P.}~\bibnamefont{Thunstr{\"o}m}},
  \bibnamefont{and} \bibinfo{author}{\bibfnamefont{K.}~\bibnamefont{Held}},
  \bibinfo{journal}{Physical Review B} \textbf{\bibinfo{volume}{93}},
  \bibinfo{pages}{165103} (\bibinfo{year}{2016}).

\bibitem[{\citenamefont{Li}(2020)}]{Li:2020aa}
\bibinfo{author}{\bibfnamefont{S.}~\bibnamefont{Li}},
  \bibinfo{journal}{Physical Review Research} \textbf{\bibinfo{volume}{2}},
  \bibinfo{pages}{013295} (\bibinfo{year}{2020}).

\end{thebibliography}

\end{document}